\pdfoutput=1
\RequirePackage{color}
\RequirePackage{ifpdf}
\documentclass{JINST}

\usepackage{graphicx}
\usepackage{subfig}
\usepackage{lineno}
\usepackage[normalem]{ulem}
\makeatletter
\AtBeginDocument{%
\DeclareRobustCommand*\subref{\@ifstar\sf@@subref\sf@subref}}
\makeatother

\title{Design and operation of LongBo: a 2 m long drift liquid argon TPC}

\author{C.~Bromberg$^a$,
B.~Carls$^b$,
D.~Edmunds$^{a}$,
A.~Hahn$^b$,
W.~Jaskierny$^b$,
H.~Jostlein$^b$,
C.~Kendziora$^b$,
S.~Lockwitz$^b$,
B.~Pahlka$^b$, 
S.~Pordes$^b$,
B.~Rebel$^b$,
D.~Shooltz$^a$,
M.~Stancari$^b$,
T.~Tope$^b$ and
T.~Yang$^b$ \\
\llap{$^a$}Michigan State University, East Lansing, Michigan 48824 USA \\
\llap{$^b$}Fermi National Accelerator Laboratory, Batavia, Illinois 60510, USA \\
}
\preprint{FERMILAB-PUB-15-104-ND}
\abstract{In this paper, we report on the design and operation of the
LongBo time projection chamber in the Liquid Argon Purity Demonstrator
cryostat. This chamber features a 2 m long drift distance. We measure 
the electron drift lifetime in the liquid argon using cosmic ray muons and the lifetime is at least 14 ms at 95\% confidence level. 
LongBo is equipped with preamplifiers mounted on the detector in the liquid argon. Of the 144 channels, 128 channels were readout by preamplifiers made with discrete circuitry and 16 channels were readout by ASIC preamplifiers. For the discrete channels,
 we measure a signal-to-noise (S/N) ratio of 30 at a drift field of 350~V/cm. The measured S/N ratio for the ASIC channels was 1.4 times larger than that measured for the discrete channels. }


\begin{document}

\section{Introduction}\label{sec:introduction}
Liquid argon time projection chambers (LArTPCs) provide excellent spatial and calorimetric resolutions for measuring the properties of neutrino interactions above a few MeV. Conventional liquid argon vessels, such as ICARUS~\cite{Amerio:2004ze}, ArgoNeuT~\cite{Anderson:2012vc} and ARGONTUBE~\cite{Ereditato:2013xaa}, are evacuated to remove water, oxygen and nitrogen contaminants present in the ambient air prior to filling with liquid argon. However, as physics requirements dictate larger cryogenic vessels to hold bigger detectors, achieving the mechanical strength required to resist the external pressure of evacuation becomes prohibitively costly. The Liquid Argon Purity Demonstrator (LAPD) \cite{Adamowski:2014daa} was an R\&D test stand at Fermilab designed to determine if electron drift lifetime adequate for large neutrino detectors could be achieved without first evacuating the cryostat. The test had two stages. In the first stage, the electron drift lifetime measured with purity monitors was greater than 6 ms without initial evacuation of the cryostat; that lifetime exceeds the value of approximately 2 ms required for future large LArTPCs with drift distances on the scale of a few meters.

After demonstrating the required electron drift lifetimes in the absence of a detector, the second stage of the test started and a TPC of 2 m drift distance, named LongBo, was installed in the central cryostat region. High liquid argon purity was achieved with the TPC in the cryostat. This paper summarizes the design and operation of the LongBo TPC.

\section{Construction of the LongBo TPC}\label{sec:tpc}
\begin{figure}[htbp]
  \begin{center}
    \includegraphics[width=0.6\textwidth]{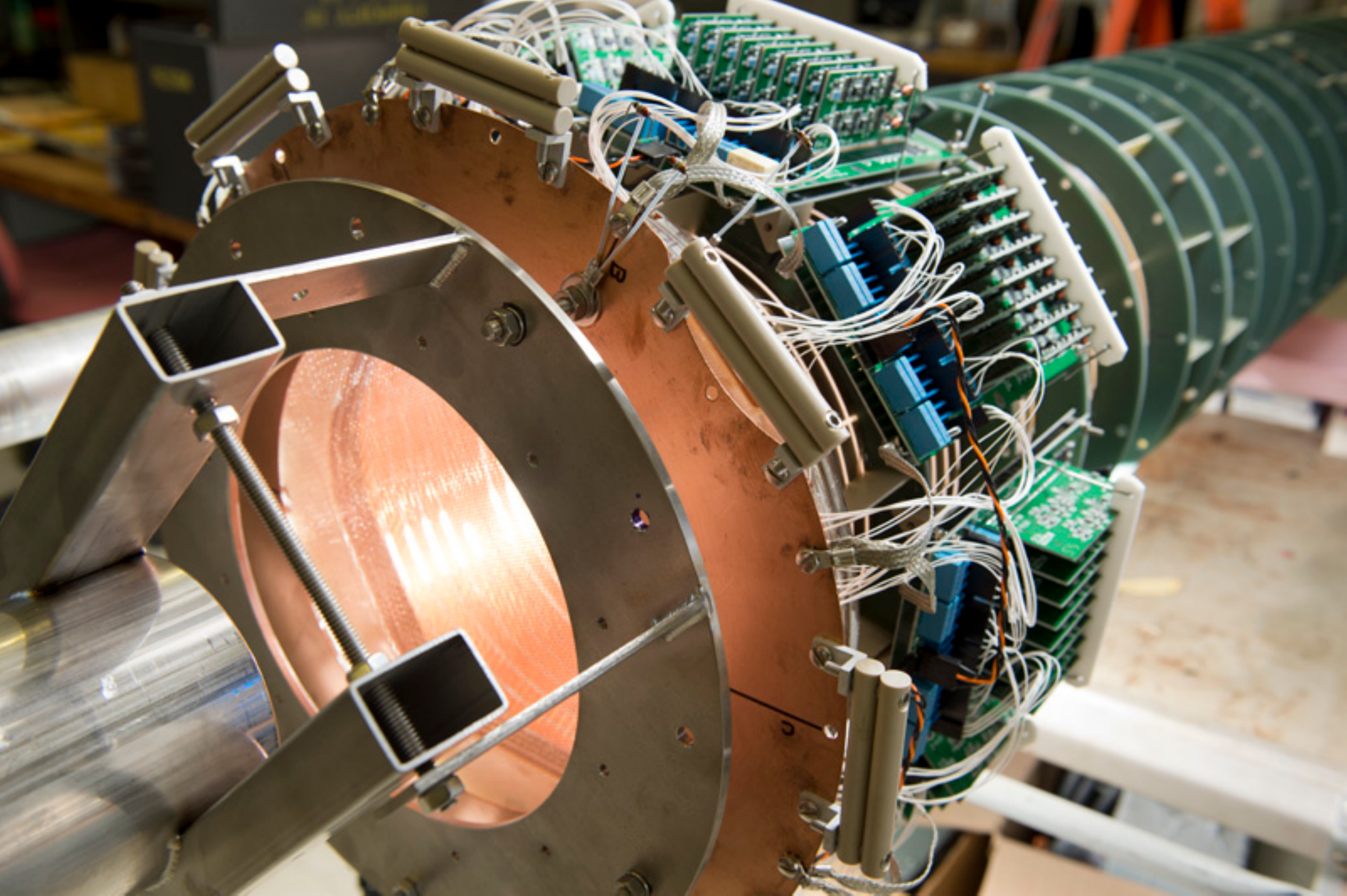}
    \caption{The wire planes and cold preamplifiers.}
\label{fig:tpchead}
\end{center}
\end{figure}

The  LongBo TPC has a cylindrical sensitive volume with a 25 cm diameter and a 2 m length. LongBo extended the drift distance of the ``Bo'' TPC that was made a few years earlier with a 50 cm length~\cite{bo}. LongBo has three sense wire planes, with wires oriented at 60 degrees from each other. There are 48 instrumented wires in each plane and the wires are 4.7 mm apart and made of 125 micron diameter beryllium copper. The wires were hand-soldered to a 3 mm thick copper-clad G10 board for each plane. The copper cladding was routed in a pad pattern on a Fermilab CNC router. A set of individual wires (shown with white insulation in figure~\ref{fig:tpchead}) were soldered to the readout boards and terminated in a connector matched to a corresponding one on the preamplifier motherboards. The preamplifiers, often referred to as ``cold preamplifiers'' since they operated in the 87 K liquid argon, were placed around the TPC just outside the field cage. The wire planes were biased from external voltage sources. The cathode was made from copper mesh, chosen to allow free argon flow along the TPC axis. The mesh was stretched inside a ring made of 19 mm diameter 3 mm thick stainless steel tube, which supports the TPC. A stainless steel tube elbow was welded to this ring and served to connect to the high voltage feed through, described in~\S\ref{sec:hv}, via a cup and spring-finger connection.

\begin{figure}[htbp]
  \begin{center}
    \includegraphics[width=0.6\textwidth]{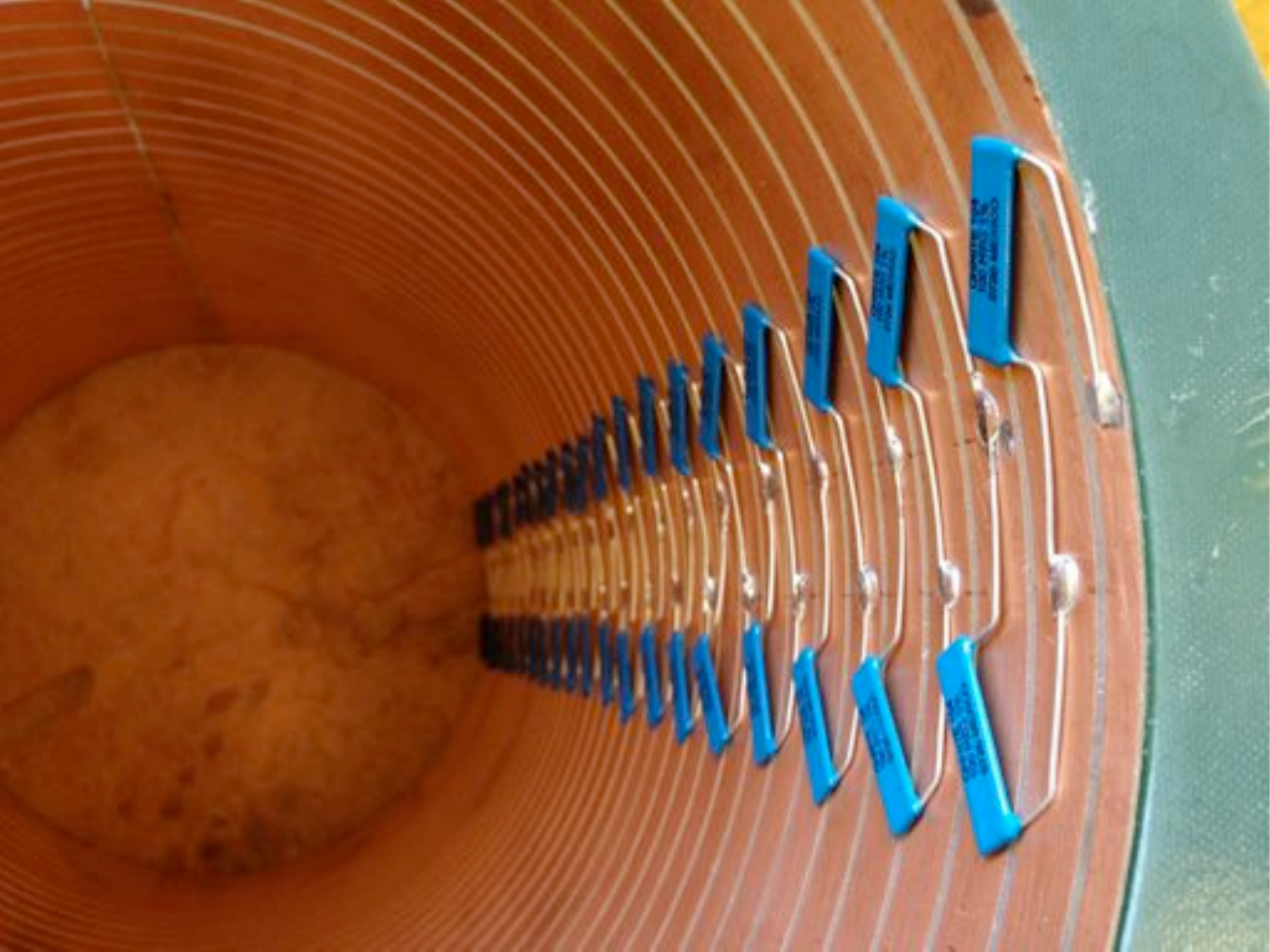}
    \caption{One chain of the resistors soldered to the inner surface of the copper sheet.}
\label{fig:resistors}
\end{center}
\end{figure}

A uniform electric drift field was provided to the original Bo TPC by a cylinder made from a copper-clad, 0.6 mm thick G10 sheet, 50 cm long by 78 cm wide. The sheet was routed to create copper strips, 18 mm wide separated by 1.5 mm. The gaps were kept narrow to provide a uniform field and minimize penetration by outside electric fields, e.g., from the feedthrough. The  sheet was rolled into a 25 cm diameter tube which was inserted into a set of 7 retaining rings routed from 4 mm thick G10. Two chains of 100 MOhm voltage divider resistors~\cite{ohmite} were soldered to the inner surface of the copper sheet. Only one chain is necessary to create the electric field; the second is to ensure continuity in case of a resistor failure. To prevent field concentration on the resistor wire leads, the resistors were installed on the inside of the cylinder where the field is uniform and small compared to the outside. The resistors were, by their location , at the same potential as the adjacent field shaping rings, avoiding significant field distortion.
Figure~\ref{fig:resistors} shows one chain of the resistors soldered to the inner surface of the copper sheet.

\begin{figure}[htbp]
  \begin{center}
    \includegraphics[width=0.6\textwidth]{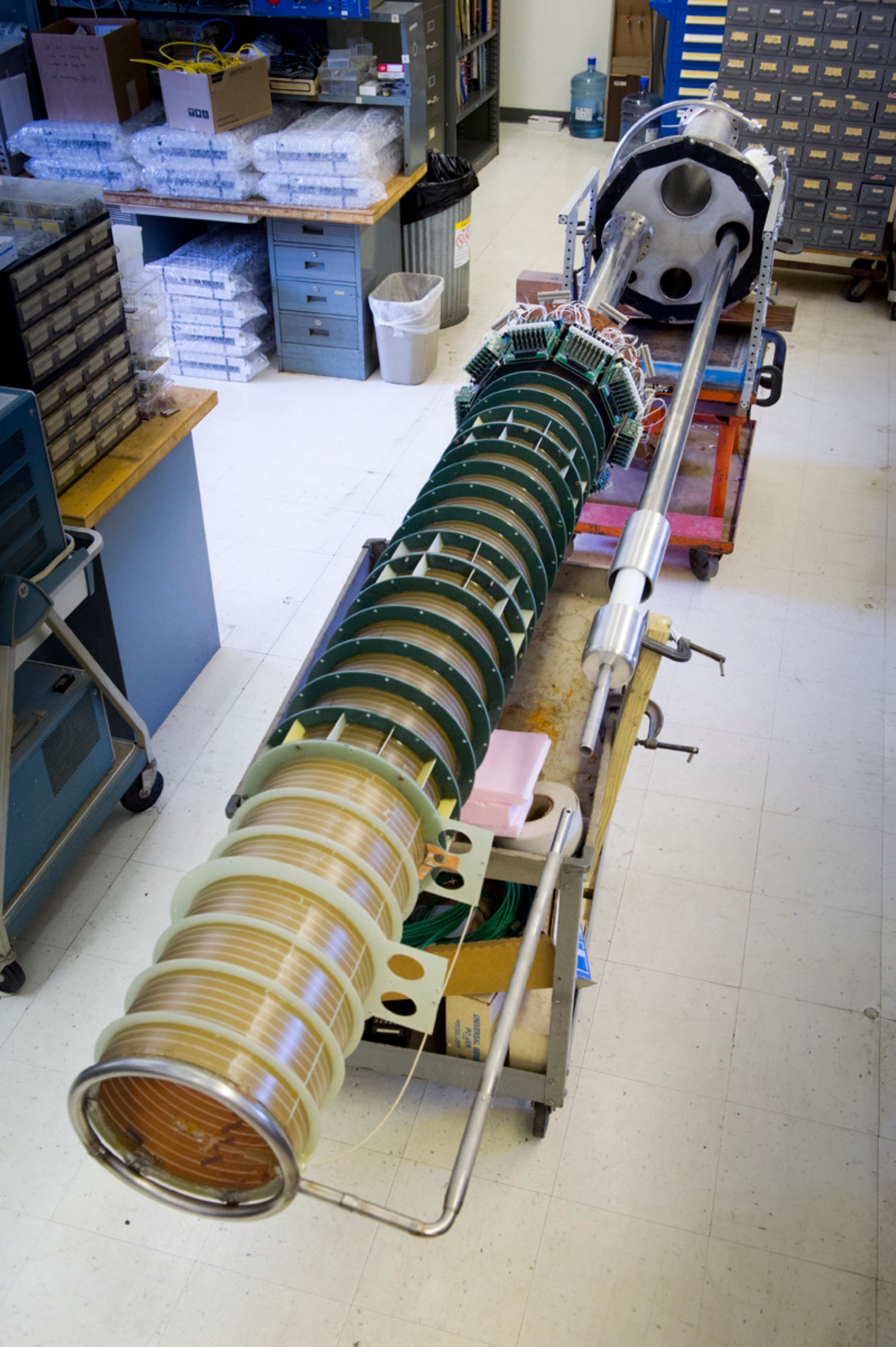}
    \caption{Assembled LongBo TPC.}
\label{fig:tpc}
\end{center}
\end{figure}

Bo was converted to LongBo by inserting  three additional  drift field sections, each 50 cm long, for a total drift length of 2 m. Each tube section was terminated, on both ends with a half-width (9mm) copper strip. At  each tube-to-tube joint the half-rings were electrically connected by two pieces of Beryllium copper spring stock soldered to the half-rings, thus ensuring smooth continuity of the electric potential. At the cathode and anode ends, the half-rings were appropriately biased to keep the drift field uniform. The drift field sections were connected together using threaded rod to make the full 2 m drift cage. Figure~\ref{fig:tpc} shows the assembled TPC with high voltage feedthrough and electronics.

\section{The High Voltage System}\label{sec:hv}
To provide the electric field to drift the signal electrons, 
a high voltage was applied to the cathode mesh on the bottom 
of the TPC. The high voltage was generated outside of the cryostat by a 
Glassman LX150N12 power supply~\cite{ref:glassman}.  The supply was controlled 
remotely by a program and in normal operation, the current is much less than 1~mA. The supply was set to trip if more than 1.1~mA 
of current was drawn in order to protect the equipment and personnel who operate the system. Before entering the cryostat, the voltage was passed through a 
``filter pot''.  
This low-pass filter was a sealed aluminum vessel with cable receptacles that 
electrically connected to a series of eight 10~M$\Omega$ resistors submerged in Diala oil.
The capacitance for the filter was supplied by the cable to the feedthrough.
The purpose of the pot was twofold:  it reduced the 
high-frequency ripple from the power supply, and it limited 
the energy to the TPC in the case of a high voltage discharge.  

\begin{figure}[htb]
  \centering
  \subfloat[]{\includegraphics[width= \textwidth]{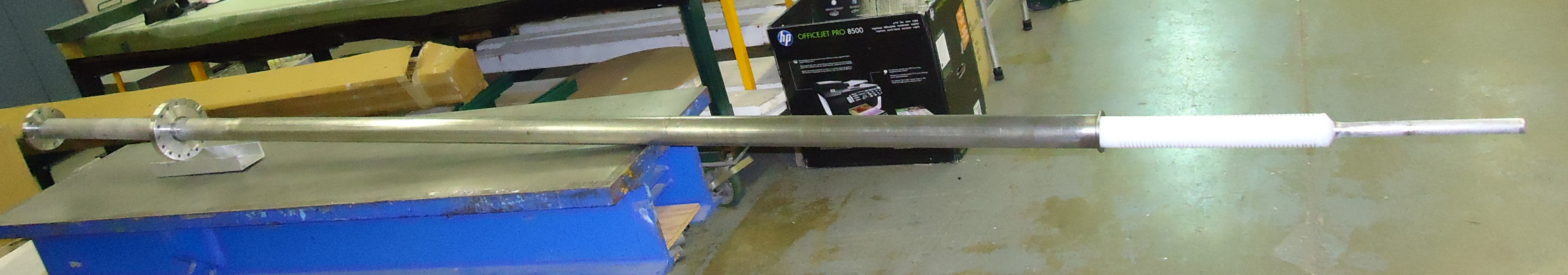}}\\
  \subfloat[]{\includegraphics[width=\textwidth]{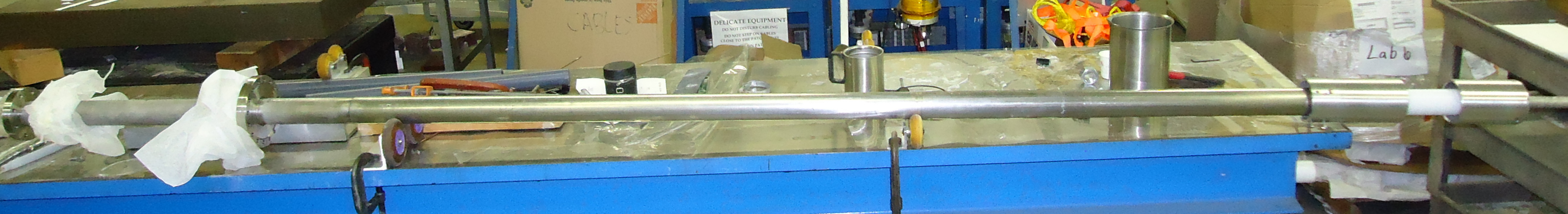}}
  \caption{Photographs of the HV feedthrough without (a) and with (b) the electric field shielding cups.}
  \label{fig:ftPict}
\end{figure}

A feedthrough transmits the high voltage into the cryostat and to the 
receptacle cup of the cathode plane.  This feedthrough is shown in figure~\ref{fig:ftPict} and was
modeled after the feedthrough used in the ICARUS experiment~\cite{Amerio:2004ze} with a 
stainless steel inner conductor insulated radially by 
ultra high molecular weight polyethylene (UHMW PE), 
and surrounded by a stainless steel ground tube.  Grooves were 
added to the exposed UHMW PE to reduce surface currents.  Initially, 
conducting shielding cups, shown in figure~\ref{fig:ftPict}b,
were installed to reduce the field along the feedthrough.  Midway through 
running, the cups were removed to determine what effect they had 
on performance.  No change was seen.  To ensure good 
electrical contact, the feedthrough has a spring tip that is inserted into 
a receptacle cup attached to the cathode plane.

\section{Trigger}\label{sec:trigger}
\begin{figure}[htbp]
  \begin{center}
    \includegraphics[width=0.8\textwidth]{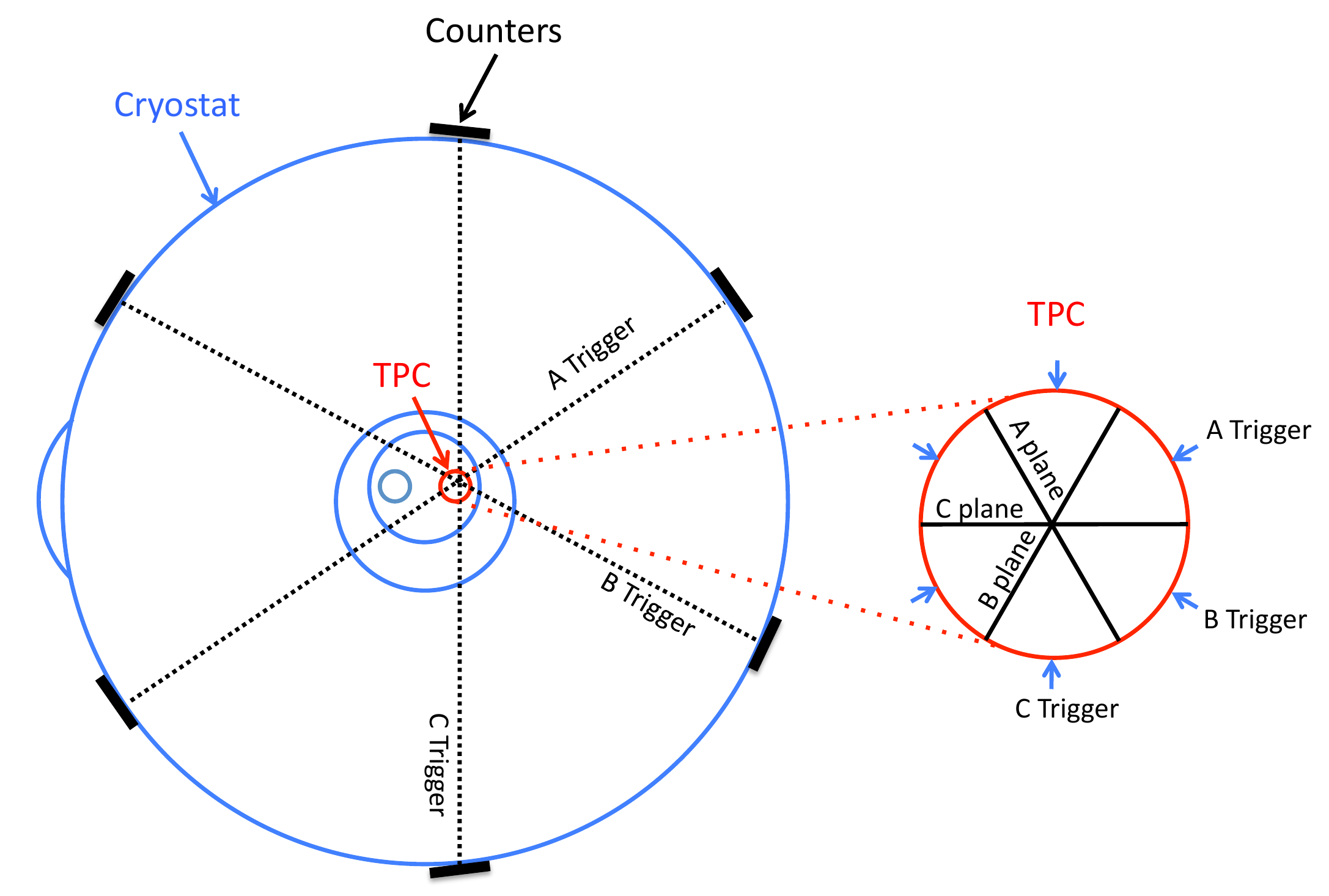}
    \caption{Location of the scintillation counters (left) and TPC wire orientations (right).}
\label{fig:triggers}
\end{center}
\end{figure}

Through-going muons were detected by three sets of 
scintillation 
counters placed around the exterior of the LAPD cryostat as illustrated in figure~\ref{fig:triggers}.  Each set
consisted of two groups of counters placed on opposite sides of the
cryostat, with the line intersecting the groups of counters passing
roughly perpendicular to the wire direction for one wire plane.
There were 8 counters (4 pairs) in each of the 6 groups, stacked in a column 
for an effective area of scintillator roughly 3 m by 0.4 m, for a total of 48 counters. One control and concentrator unit (CCU) controls 48 counters setting the gain of the  photomultiplier tube (PMT) and a discriminator threshold~\cite{Bromberg:2001gd}.
A coincidence was required between counters on opposite side of the
cryostat, and the track angle coverage is shown in
\S~\ref{sec:results}.

\section{Electronics}
\subsection{Cold Electronics}\label{sec:electronics}
The LArTPC technology was pioneered by the ICARUS collaboration using warm electronics with a dual-jFET charge integrating front-end preamplifier. For each wire of the ICARUS detector the feed-through passed DC bias-voltage into the cryostat and passed the TPC signals back to the electronics. Early development in the US used a similar warm preamplifier on a 3-plane LArTPC detector, Bo~\cite{bo}, built by Fermilab and used a 32-channel ADF2 digitizer built by Michigan State University (MSU) for the D0 experiment. At the same time the ArgoNeuT \cite{Anderson:2012vc} LArTPC was built and used the same warm preamplifier design readout.

\begin{figure}[htp]
\centering
\subfloat[Cold motherboard.] 
{
    \label{fig:cmb:a}
    \includegraphics[width=.45\textwidth]{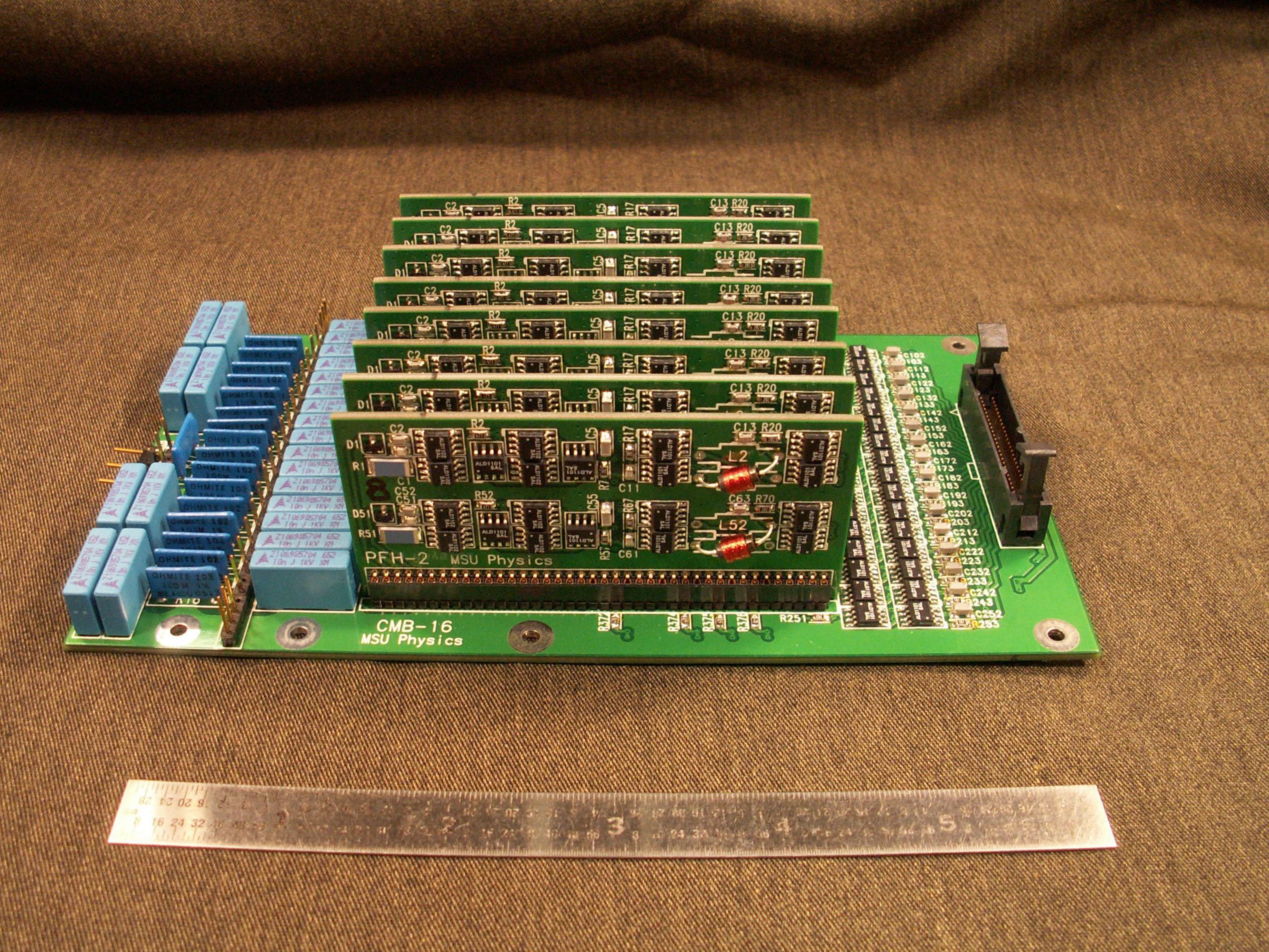}
}
\subfloat[TPC with preamp-filters mounted.] 
{
    \label{fig:cmb:b}
    \includegraphics[width=.45\textwidth]{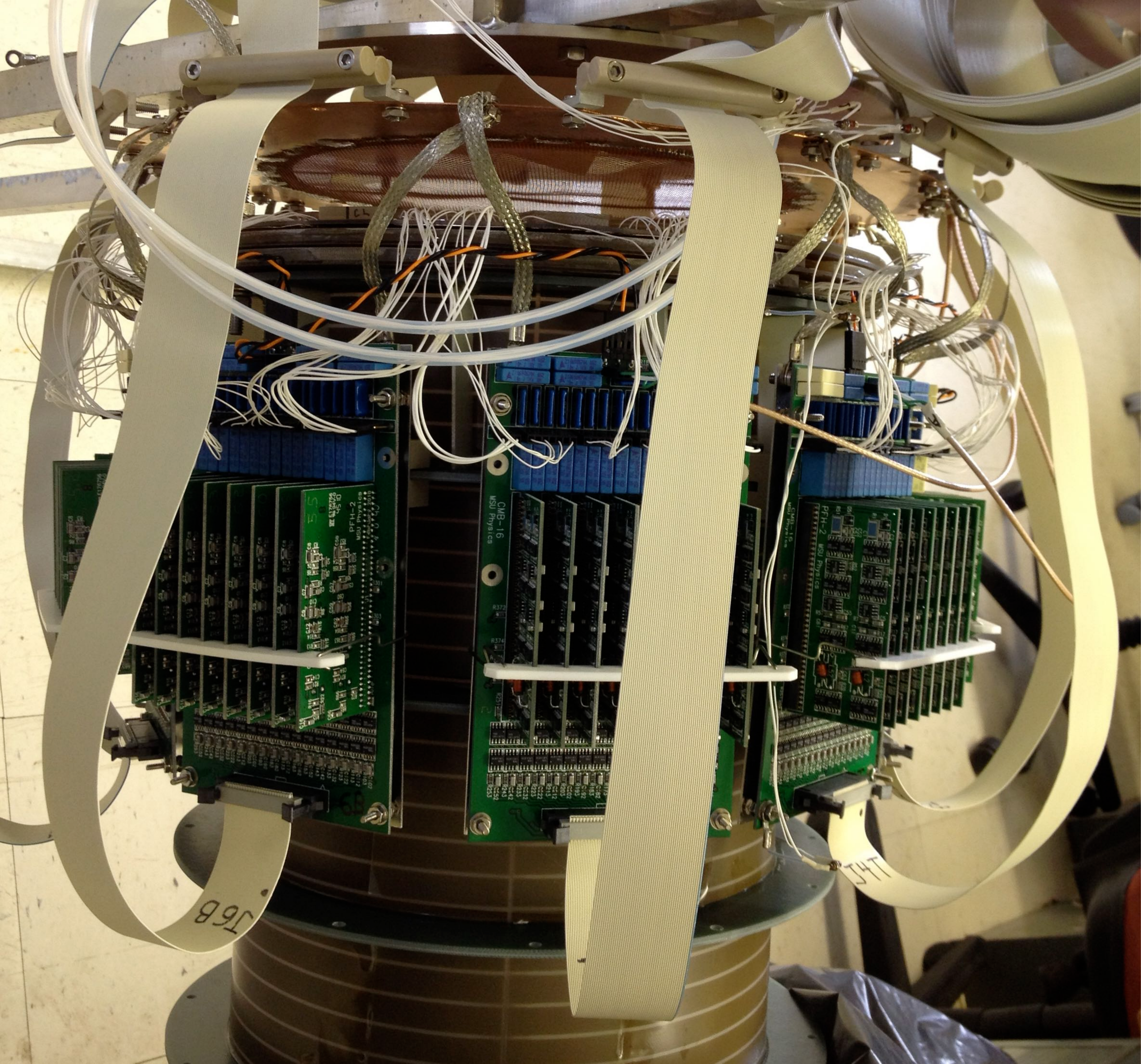}
}
\caption{Left: Cold motherboard with 8 preamp-filter cards. Right: TPC with preamp-filters mounted near the wire planes. The flat Polyolefin cables connect the output of the cold CMB cards to the cold side of the signal feedthrough card.}
\label{fig:cmb}
\end{figure}

The development of cold electronics for a LArTPC was driven by the recognition of the freedom in TPC design provided by being able to place the preamplifiers other than at the top of the wire planes and by the desire to avoid the noise generated by the lengthy cables required for large detectors. Additionally, at the 87 K temperature of liquid argon there is less intrinsic noise in a MOSFET preamp than in the best warm preamplifier. 
72 dual-channel hybrid preamplifiers were designed and assembled at MSU. The channels were designed for operation at 87 K, using a CMOS transistor chip front-end configured as a charge integrator, followed by a combination of high and low pass filters characterized by a step-function response with a 2.4 $\mu$s peaking time. The cold preamp-filter cards were first used by the Bo TPC to replace the original warm preamp-filters and then used by the LongBo TPC. Cold motherboards (CMB) each carrying 8 dual-channel hybrids, were mounted on brackets, as shown in figure~\ref{fig:cmb}, close to the wire planes. The 9 CMB boards read out 144 channels in total. Short insulated wires connected each CMB to a TPC wire plane. Flat Polyolefin cables carried the output of each CMB to a warm feedthrough. Cable drivers on the outside of the feedthrough drove signals to ADF2 digitizers sampling at 2 MHz.

A MOSFET ASIC preamp-filter for use at cryogenic temperatures was designed by BNL~\cite{Thorn:2012zsa} for DUNE~\cite{Adams:2013qkq}. The ASIC is also the readout of the TPC in the MicroBooNE experiment~\cite{microboone}. A version (V4) of this 16-channel ASIC became available in 2012 and was implemented for the LongBo TPC. One 16-channel preamp-filter card was modified to use a section of the MicroBooNE motherboard that serviced a single ASIC as a mezzanine board. The input/output connectors, bias-voltage distribution and decoupling capacitors of the card were retained. An ASIC control box was built which set the gain and peaking time parameters of the ASIC. The ASIC modified preamp-filter card was installed in place of one induction plane card with the signals processed by the same digitizing electronics.

\subsection{Signal to Noise Ratio}\label{sec:ston}
The signal to noise ratio is an important metric of performance for a 
liquid argon detector.  The absolute numerical value depends upon wire
spacing, wire capacitance, readout electronics and drift field strength.
Because 16 channels of the LongBo readout electronics were
instrumented with the BNL ASIC preamplifier, a direct performance comparison can
be made between the ASIC version of the preamplifier and the discrete
CMOS version used for the remaining channels.

\begin{figure}[htbp]
\centerline{\includegraphics[scale=0.5]{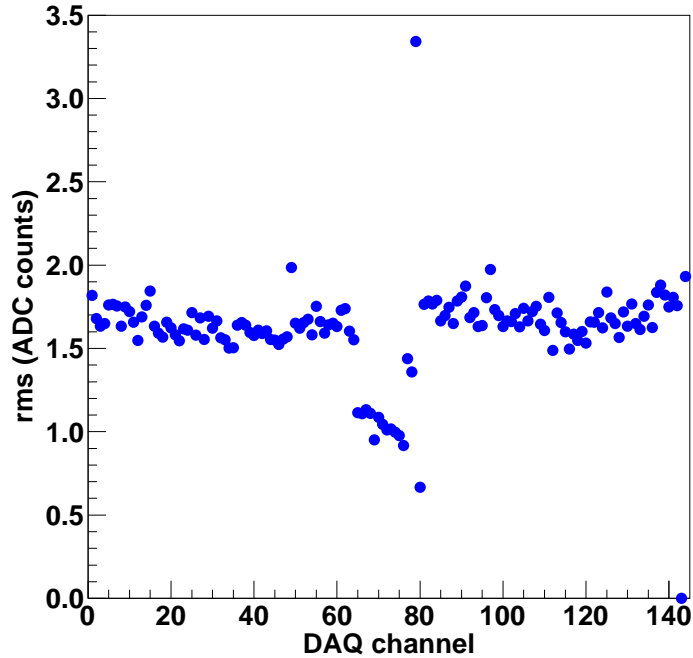}}
\caption{Measured noise as a function of DAQ channel number.  Channels
1-48 are the first induction plane, channels 49-96 are the second
induction plane and channels 97-144 are the collection plane.
Channels 65-80 are instrumented with the BNL ASIC preamplifier, all other
channels are instrumented with the MSU discrete CMOS preamplifier. After the first HV breakdown, four channels of the ASIC readout (channels 77-80) exhibited more noise than before and it was no longer possible to change their parameter settings. All of the MSU discrete CMOS channels survived the HV breakdowns.
}
\label{fig:noise}
\end{figure}

To measure the noise for each channel,
a special run of 200 triggers was taken with a random trigger and no
drift field.  For each channel, the measured ADC counts for all 200
triggers were combined and the distribution fit to a Gaussian function.
The width of the best fit Gaussian is reported as the noise in 
figure~\ref{fig:noise}.
The measured noise in figure~\ref{fig:noise} can also be determined
from regular
data runs with an external muon trigger and nominal drift field if
signals from cosmic ray muons are removed from the acquired waveforms.

The signal strength was extracted from straight muon tracks acquired
with the cosmic ray muon trigger, which is described in \S~\ref{sec:trigger}.
The signal to noise 
ratio of the discrete electronics chain was determined from signals
on the collection plane wires.  
Clean tracks were selected 
from the first 200 triggers in each run considered
using the event display. Only tracks whose projection 
onto the wire plane under study was nearly
perpendicular to wire direction and centered on the wire plane
were selected.
The signal strength is recorded as 
the maximum ADC count in the waveform.
The outermost two wires on either edge of the wire
plane were excluded from the analysis as a fiducial volume cut
intended to reject hits close to the field cage where the drift field may not be uniform and the pulse heights are noticeably reduced.
In addition, wires exhibiting pulse heights greater than 1.5 times the
average pulse height for the track were interpreted as delta rays and
excluded.  Finally, wires with more than one hit for a given trigger
were excluded.

In the absence of diffusion, the length of the track 
segment that contributes to 
the signal on a particular collection plane wire is 
$\Delta s\equiv\frac{\Delta w}{\sin(\theta)\cos(\phi)}$, where $\Delta w=4.7$~mm is 
the wire spacing and 
$\theta$ and $\phi$ are the reconstructed track angles defined
in \S~\ref{sec:results}.  The individual measured signal strengths were
reduced by a factor of $\sin(\theta)\cos(\phi)$ before taking 
the average of all {\em good} collection wire values 
from the 10-12 event sample 
to obtain the value of the signal strength $S=50.3$ ADC counts 
and a signal to noise ratio of $50.3/1.69 \sim 30$
for the discrete CMOS electronics chain.

The signal to noise
ratio provides a good comparison of the performance of
the different amplifier flavors. The ASIC preamplifier
is connected to the center 16 wires of the
middle (induction) wire plane, so induction plane signals must be
compared.   For these bipolar signals, the signal strength 
used is the difference between the pedestal and the 
minimum ADC count of the waveform, or the depth of the dip in the
bipolar signal shape.
 Tracks whose projection onto
the middle wire plane is perpendicular to the wire direction 
were selected ($\phi \sim 60^{\circ}$)  and the extracted signal strengths
were normalized by the
factor $\sin(\theta)\cos(\phi-60^{\circ})$.
The ratio of the average signal strength for ASIC and discrete channels
is reported in table~\ref{table:result} for different values of the
shaping time constant.  Both the measured noise
and the signal to noise ratio increase with the shaping time constant.
\begin{table}[htp]
\centering
\caption{Comparison of signal and noise for discrete CMOS preamplifier
  and the ASIC preamplifier for different settings of the shaping time
  constant of the ASIC. The discrete CMOS preamplifiers have a fixed
  shaping time of 2~$\mu$s. The gain setting of the ASIC
  was 25 mV/fC for all measurements tabulated. }
\begin{tabular}{cccccc}
\hline
Shaping time ($\mu$s) & $S$&$S_{discrete}$ & $N$&$N_{discrete}$  &
$(S/N)/(S/N)_{discrete}$
\\ \hline
 0.5  & 10.1 & 36.1 &  0.55 & 1.69 & 0.9 \\ \hline 
 1.0  & 19.3 & 35.1 &  0.75 & 1.69 & 1.2 \\ \hline 
 2.0  & 33.6 & 35.5 &  1.10 & 1.69 & 1.4 \\ \hline 
 3.0  & 40.3 & 35.8 &  1.35 & 1.69 & 1.4 \\ \hline 
\end{tabular}
\label{table:result}
\end{table}

\begin{figure}[htbp]
\centerline{\includegraphics[width=0.8\textwidth]{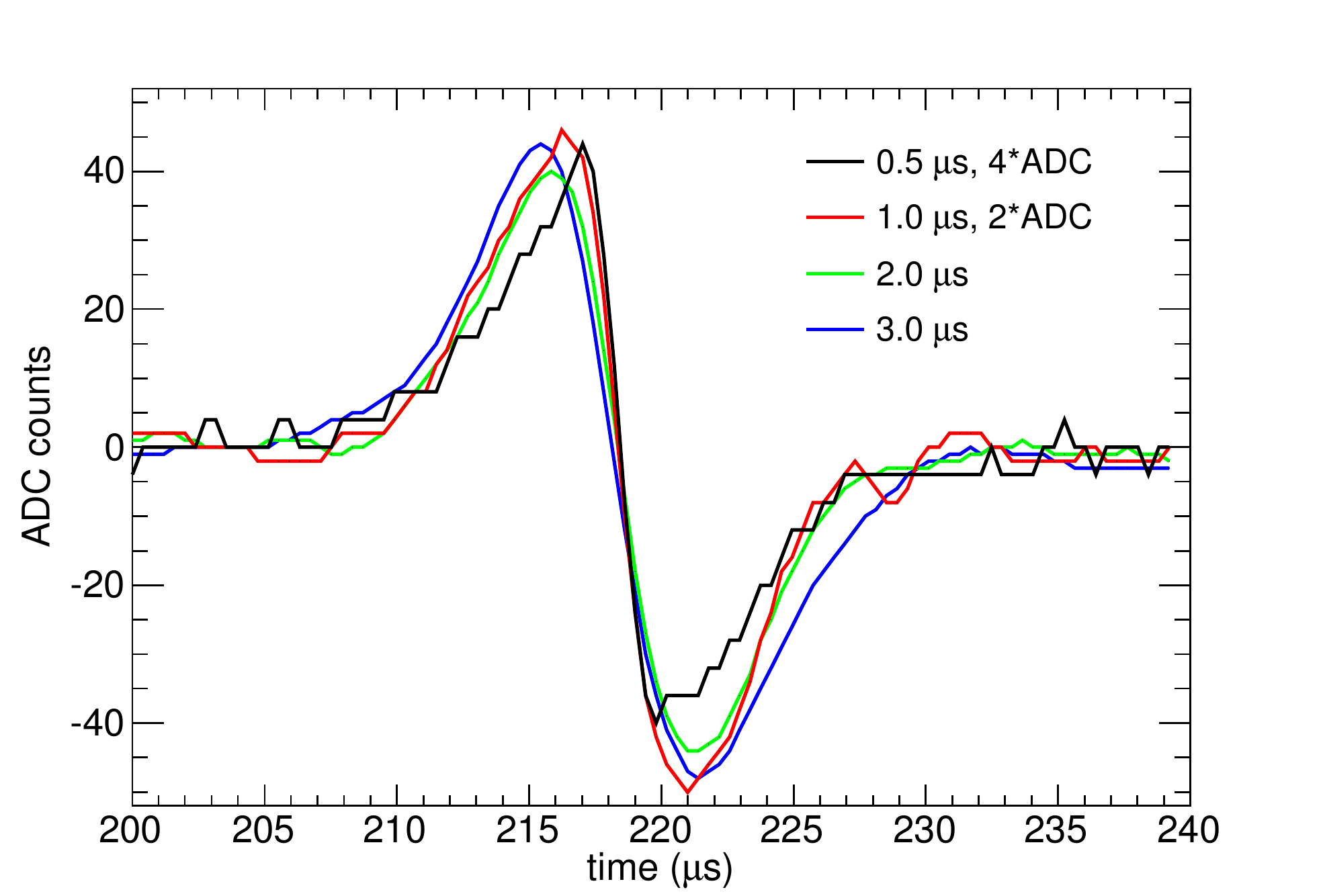}}
\caption{Raw signal shapes for induction plane signals with different
  settings of the ASIC shaping time constant are shown. 
The horizontal axis has an arbitrary zero so that the pulses 
are shown with the zero crossing point artifically aligned.
The tracks from which these signals are taken form an angle of roughly
20~degrees with the plane of the wires, or $\theta \sim 70$.  The angle
between the  projection
of the track onto the wire plane and the wires themselves is $\sim80$~degrees.}
\label{fig:compare}
\end{figure}

Figure~\ref{fig:compare} shows the shape of the amplified signal 
on a single wire 
for different shaping times, with the the signals scaled to 
have the same peak height to facilitate comparison. 
There is little if any difference in the pulse shapes - the shape is 
dominated by the width of the pulses before amplification.  Changing
the shaping time constant changes the overall gain of the amplifier.

\section{Operation}\label{sec:operation}
\begin{figure}[htbp]
  \begin{center}
    \includegraphics[width=0.5\textwidth]{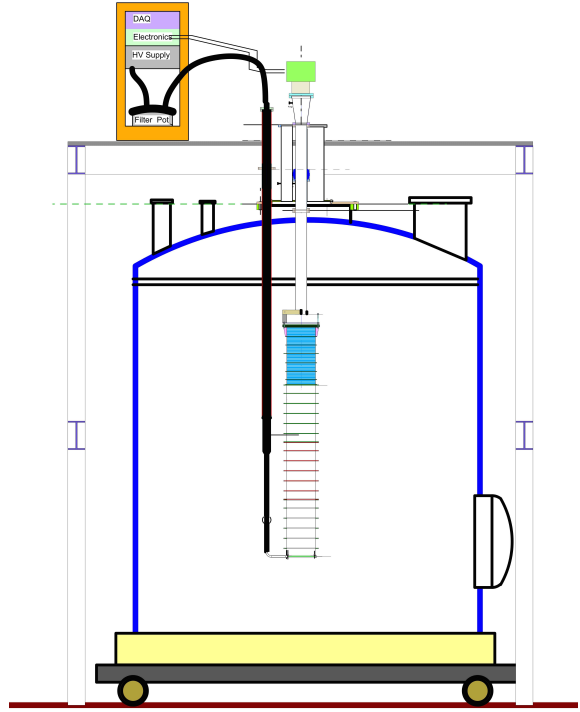}
    \caption{The TPC and high voltage feed through inside the LAPD cryostat.}
\label{fig:tpcintank}
\end{center}
\end{figure}

The LAPD cryostat is an industrial low pressure storage tank. The cryostat has a flat bottom, cylindrical sides, and a dished head. The cryostat diameter is 3.0 m and the cylindrical walls have a 3.0 m height. The LongBo TPC was inserted into the LAPD cryostat in October, 2012. Figure~\ref{fig:tpcintank} shows the positions of the TPC and the high voltage feed through inside the LAPD cryostat. 
After purging the cryostat with argon gas and then filtering the argon gas for several days, as described in~\cite{Adamowski:2014daa}, the cryostat was filled with liquid argon. The high voltage was raised after the electron drift lifetime passed 0.3 ms. The TPC data taking lasted for 8 months. 

An issue we encountered in the LongBo run was that the high voltage was unstable. The cause of the instability is unknown, but was present during the entire data-taking period. This instability forced the running of the TPC at a lower high voltage than desired, ultimately allowing a maximum of -75 kV, with occasional trips, instead of the desired -100 kV for a 2 m drift TPC. Even with the lower-than-desired high voltage, several hundred thousand clear cosmic ray muon tracks were recorded. These data have been used to study the electronics performance and liquid argon purity.

\section{Results}
\subsection{Measurement of Electron Attenuation Using Cosmic Ray Muons}\label{sec:results}
When cosmic ray muons pass through the liquid argon, they deposit energy resulting in ionization and scintillation. The ionization electrons that escape recombination are drifted by the electric field and collected by the wire planes. The variation of the energy deposition along the muon track is small for muons with sufficient energy to produce triggers. By examining the signal recorded by each wire as a function of electron drift time, one can measure the attenuation of ionization electrons along the drift distance and determine the electron drift lifetime. This method was used by ArgoNeuT~\cite{Anderson:2012vc} and ICARUS~\cite{Antonello:2014eha} to derive an electron drift lifetime.

We analyzed cosmic ray muon data taken during one full cycle of LAPD running, from when the liquid argon started recirculating through the filters to when it stopped. The high voltage we applied to the cathode was -70 kV during this period, which produced an electric field of 350 V/cm in the TPC volume. In this section, we present results on the measurements of electron attenuation using these data.

\begin{figure}[htp]
\centering
\subfloat[Raw data] 
{
    \label{fig:reco:a}
    \includegraphics[width=.59\textwidth]{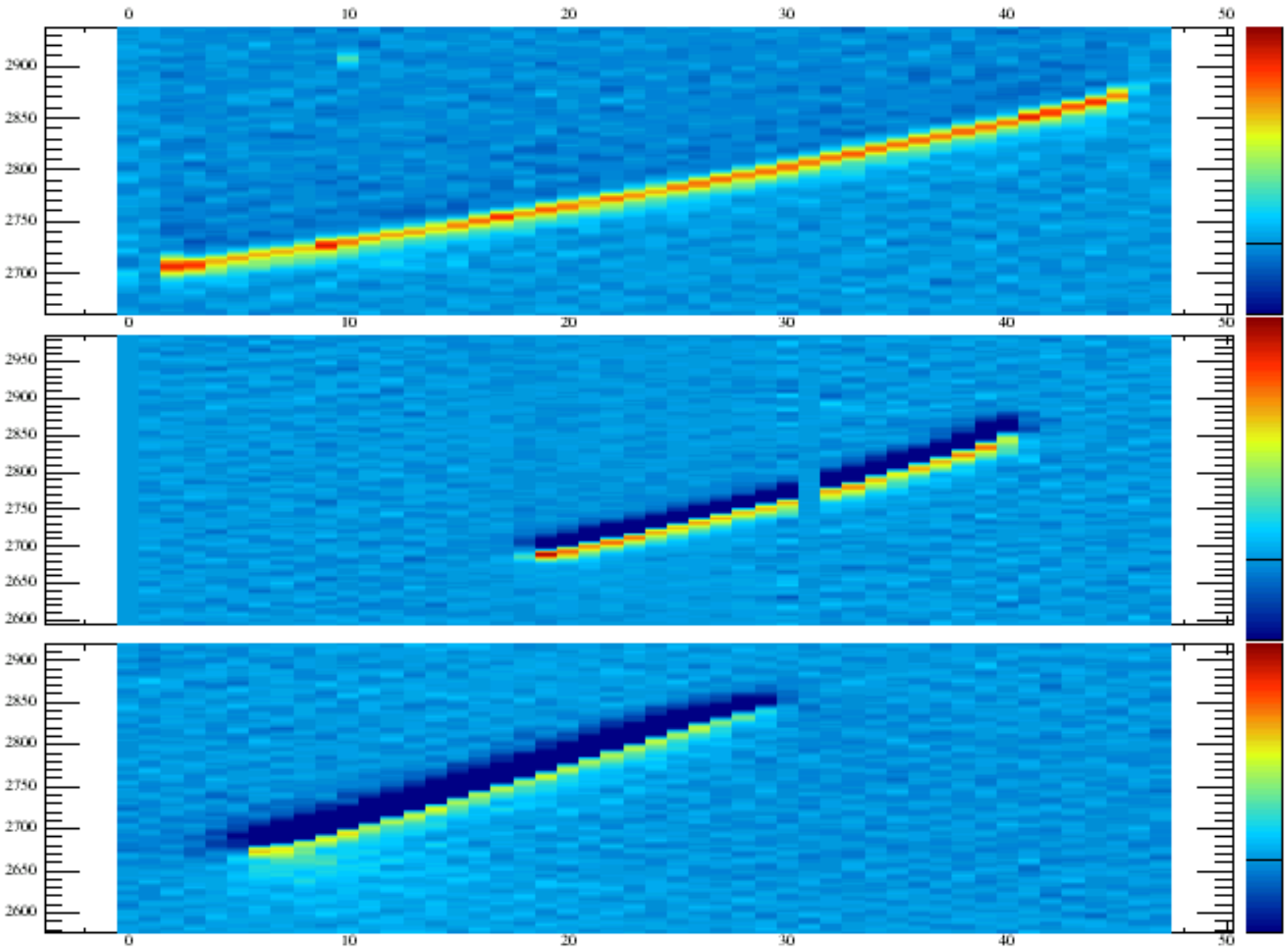}
}
\subfloat[Reconstructed track] 
{
    \label{fig:reco:b}
    \includegraphics[width=.31\textwidth]{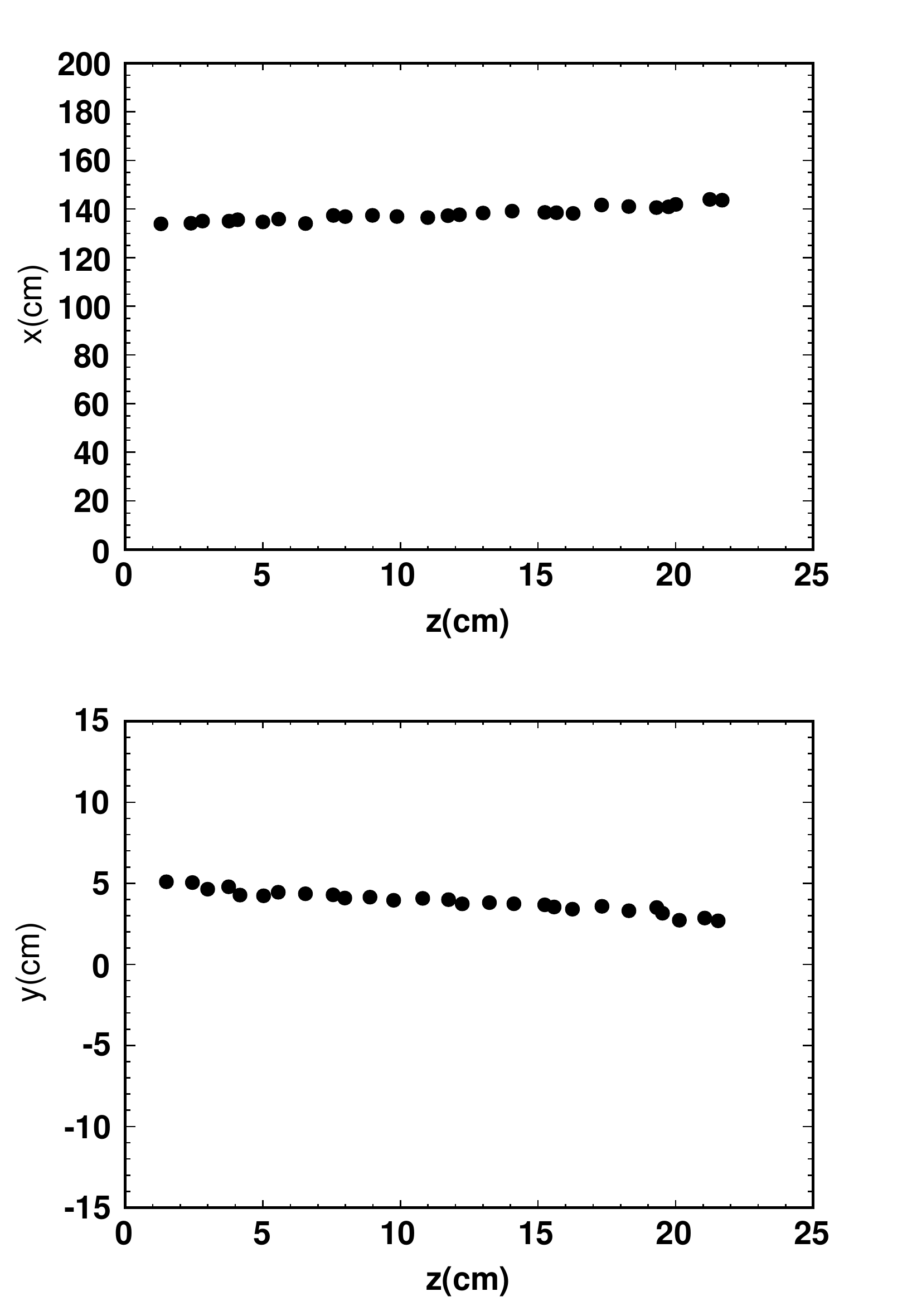}
}
\caption{Run 293 Event 2225. (a) Drift time versus wire ID for raw wire signal from 3 wire planes. Red indicates positive signals and blue indicates negative signals. (b) Reconstructed track points in the x-z and y-z projections.}
\label{fig:reco} 
\end{figure}

There are 491 966 triggered events during this run period. We use the {\sc larsoft} software package~\cite{larsoft} to reconstruct cosmic ray muon events. The automated reconstruction first converts the raw signal from each wire to a unipolar pulse by a deconvolution, which accounts for the electronics response to the induced currents, and then finds hits by fitting a Gaussian to the resulting pulse. The hits from each plane of the TPC are grouped into clusters in drift time and wire number. Three dimensional tracks are constructed from pairs of line-like clusters in each plane. There are 274 491 events with at least one reconstructed track. Figure~\ref{fig:reco} shows one example event where a muon track is visible. 

\begin{figure}[htp]
\centering
\subfloat[Track $\theta$] 
{
    \label{fig:angle:theta}
    \includegraphics[width=.45\textwidth]{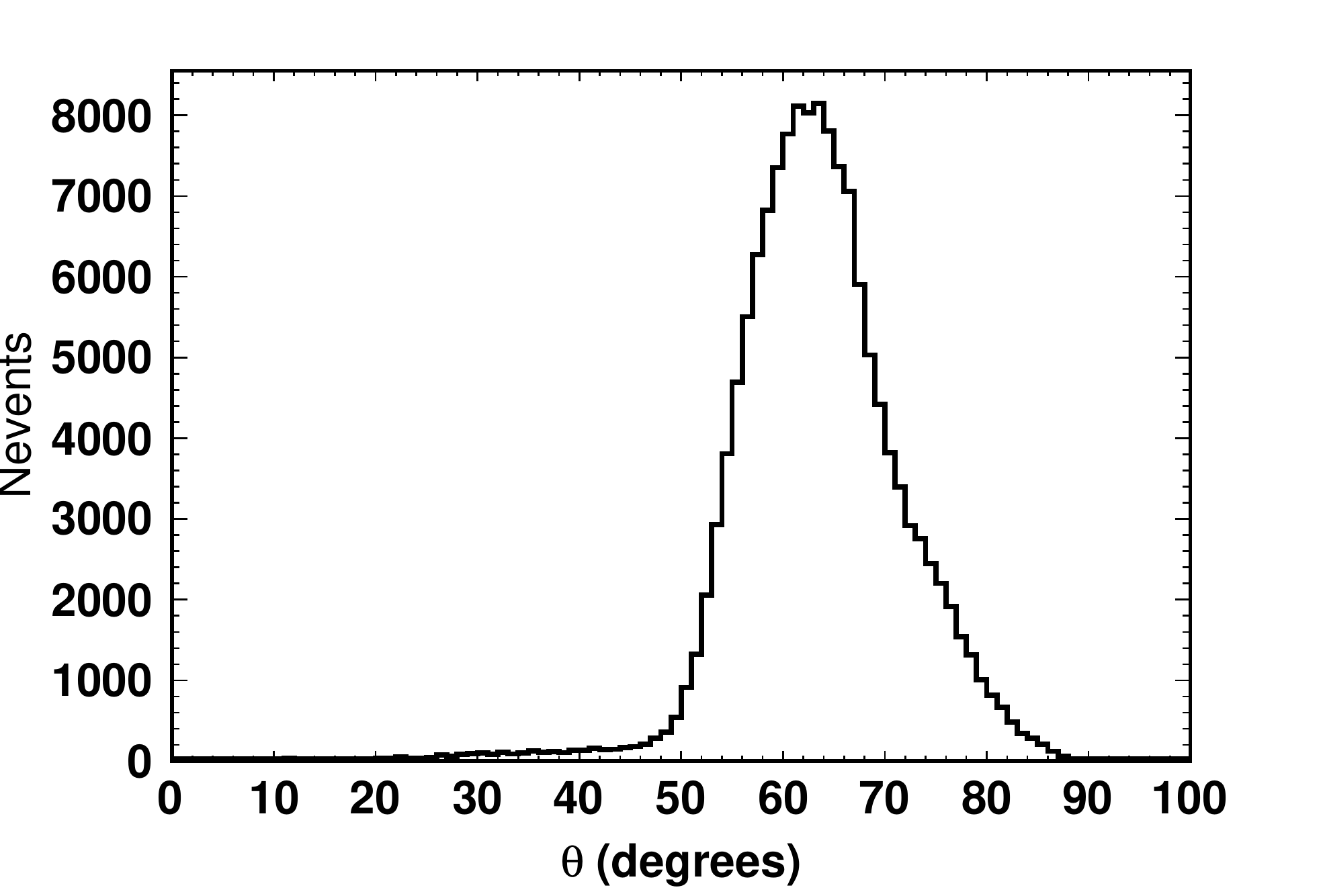}
}
\subfloat[Track $\phi$] 
{
    \label{fig:angle:phi}
    \includegraphics[width=.45\textwidth]{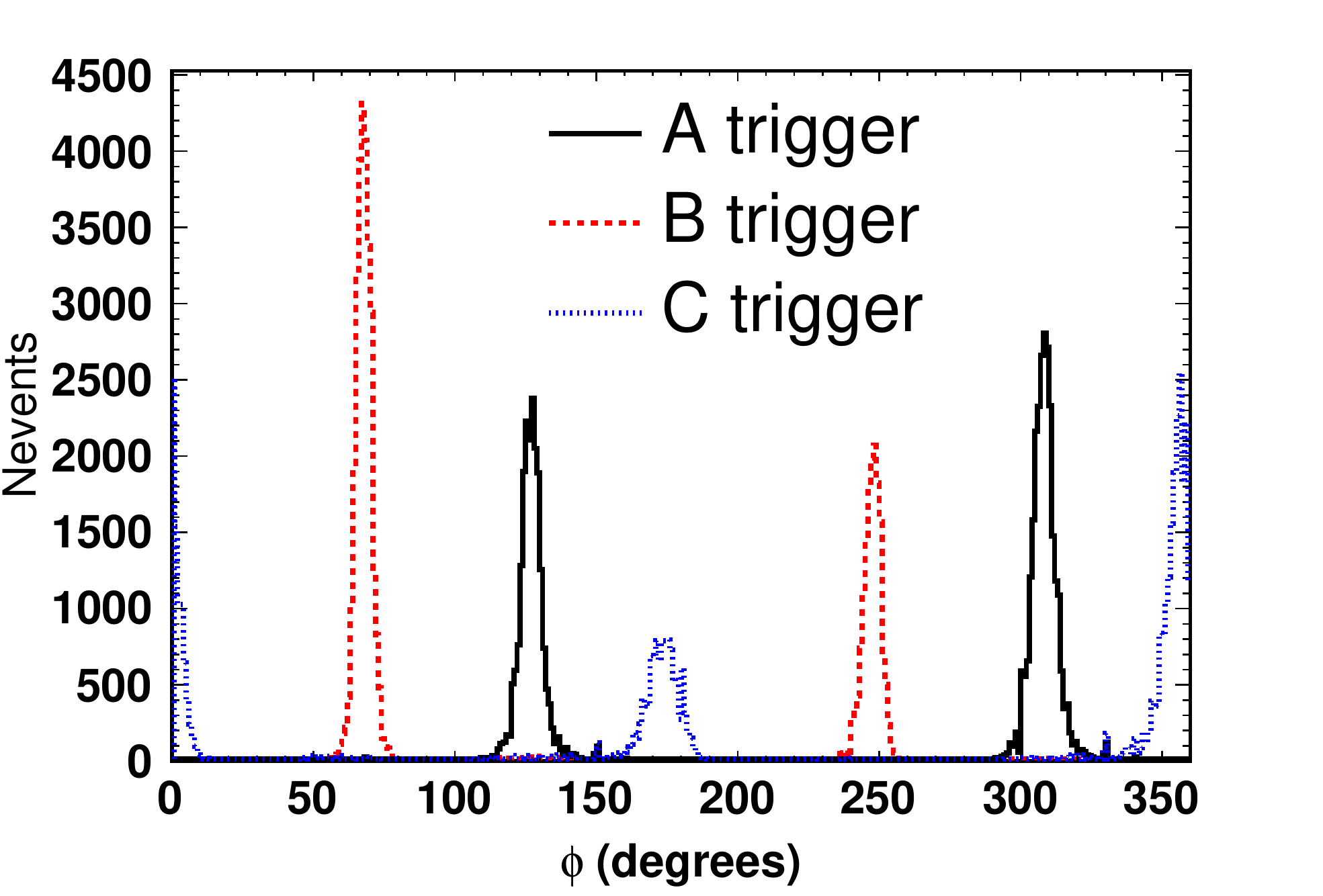}
}
\caption{Reconstructed track angles: (a) $\theta$ - with respect to the vertical direction, while $\theta = 0$ is at the zenith; (b) $\phi$, events taken with different triggers have distinct $\phi$ distributions. }
\label{fig:angle} 
\end{figure}

\begin{figure}[htbp]
  \begin{center}
    \includegraphics[width=0.8\textwidth]{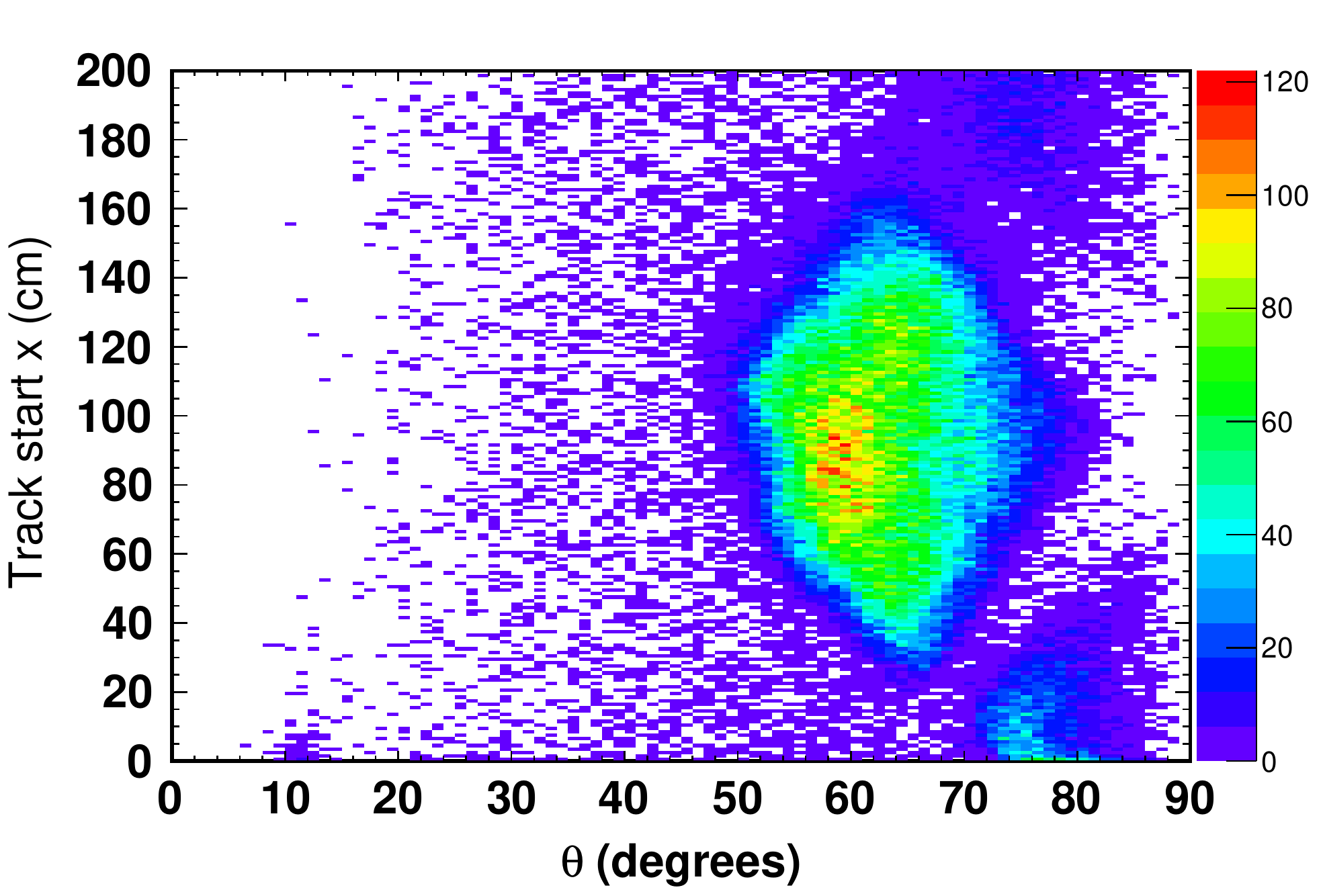}
    \caption{Reconstructed track start point in the drift direction ($x$) versus track angle ($\theta$). The muons normally enter the TPC in the middle, but they can be close to the wire planes and cathode. }
\label{fig:trkxtheta}
\end{center}
\end{figure}

In this analysis we consider the 223 194 events with exactly one reconstructed track that has at least 5 reconstructed 3-dimensional points along the track trajectory. We do not use events with multiple muon tracks because we cannot determine the track start time, $t_{0}$, for the individual tracks. Figure~\ref{fig:angle} shows the angular distributions of the reconstructed tracks. The track angle is determined by the reconstructed track start and end points. The zenith angle $\theta$ is peaked around $60^{\circ}$ and the azimuthal angle $\phi$ is determined by the trigger counter locations. Figure~\ref{fig:trkxtheta} shows the reconstructed track start point $x$, in the drift direction, versus the $\theta$ angle; $x = 0$ is near the wires. The distribution in height is concentrated in the middle of the TPC with adequate statistics at the extreme ends of the chamber. 

The determination of the electron drift lifetime is performed using the collection plane hits on the track. We first select events with well reconstructed and clean tracks, where
\begin{enumerate}
\item the RMS distance of reconstructed 3D points with respect to a straight line determined by track start and end points is less than 6.3 mm;
\item the number of hits that are not associated with the reconstructed track is less than 11;
\item the track length is greater than 15 cm;
\item the track length $\times\cos\phi$, the projection along the collection plane trigger direction, is greater than 10 cm;
\item $\theta$ is between $50^{\circ}$ and $70^{\circ}$.
\end{enumerate}
The selection was chosen by scanning the distributions to maximize the good track fraction. We then select all hits on the collection plane associated with the tracks. We require only one hit on each wire. If there is one hit on each of three contiguous wires that has $dQ/ds > 2000$ ADC/cm, while $dQ/ds$ is the charge deposited per unit length along the particle trajectory, we do not use those three hits. Both requirements are meant to remove delta rays.

For each selected hit, we look at the raw wire signal for that hit in the region [$t-3\sigma_{hit}$, $t+3\sigma_{hit}$], where $t$ is the reconstructed hit time and $\sigma_{hit}$ is the reconstructed hit width. We first find the peak of the raw counts in the ADC, we then sum all the counts above a threshold of 10\% of the peak to obtain the charge of the hit. The threshold is chosen based on the measured signal to noise ratio. By using the raw wire signal we remove uncertainties introduced by signal shaping in the hit reconstruction. We then divide the hit charge by the track pitch, which is defined as $\Delta s$ in \S~\ref{sec:ston}, to obtain $dQ/ds$ for the hit. We look at the $dQ/ds$ distributions in bins of 51 $\mu s$ drift time and fit a Landau convoluted with Gaussian function to the individual $dQ/ds$ distributions. We plot the the most probable value (MPV) from the Landau fit as a function of drift time and then we fit an exponential function to the distribution:
\begin{equation}
dQ/ds = dQ/ds_{0}e^{-at_{drift}},
\end{equation}
where $dQ/ds_{0} \equiv dQ/ds(t_{drift}=0)$ is the charge deposited by the muon, $t_{drift}$ is the drift time and $a\equiv1/\tau$ is the attenuation constant while $\tau$ is the electron drift lifetime.

\begin{figure}[htp]
\centering
\includegraphics[width=.45\textwidth]{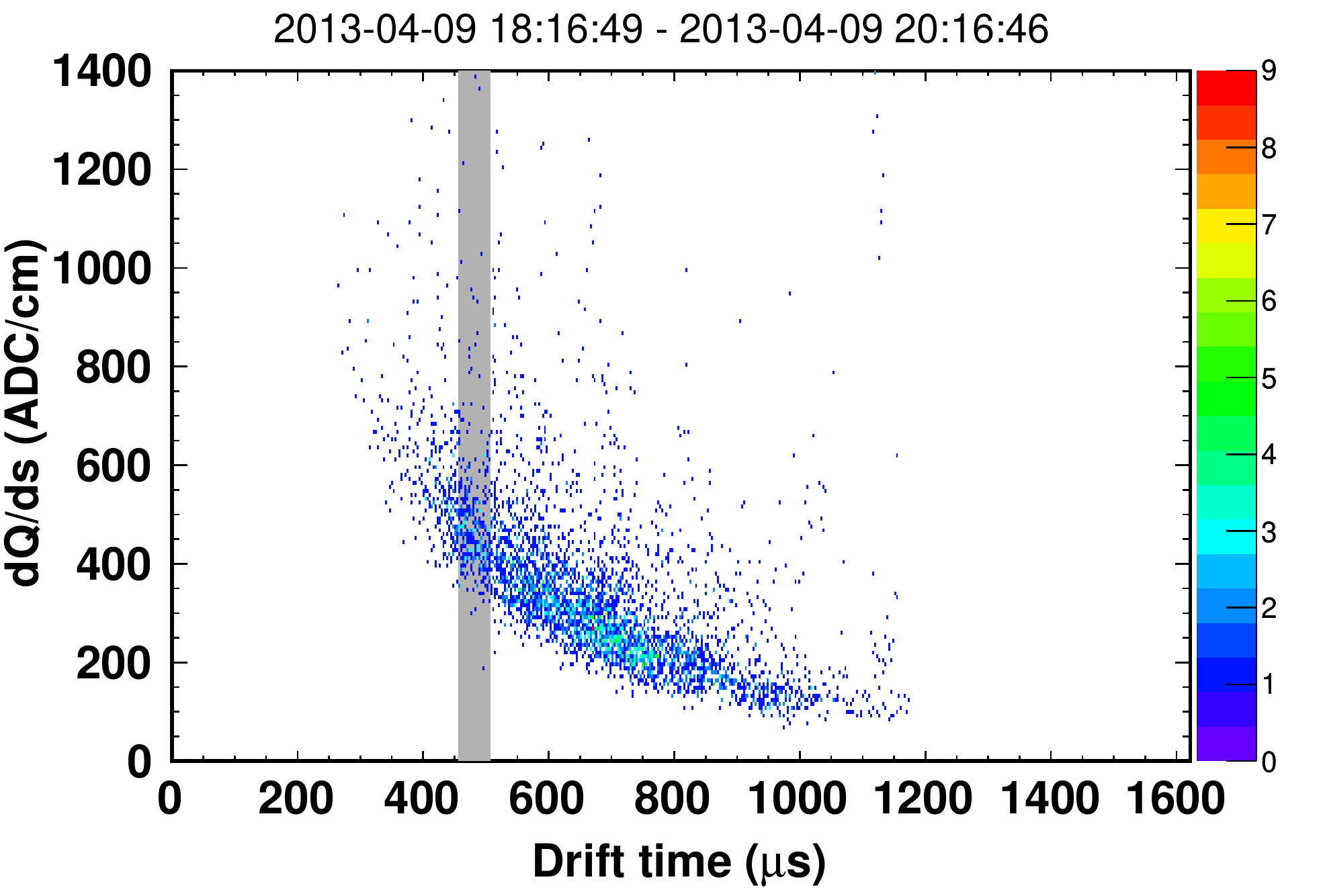}
\includegraphics[width=.45\textwidth]{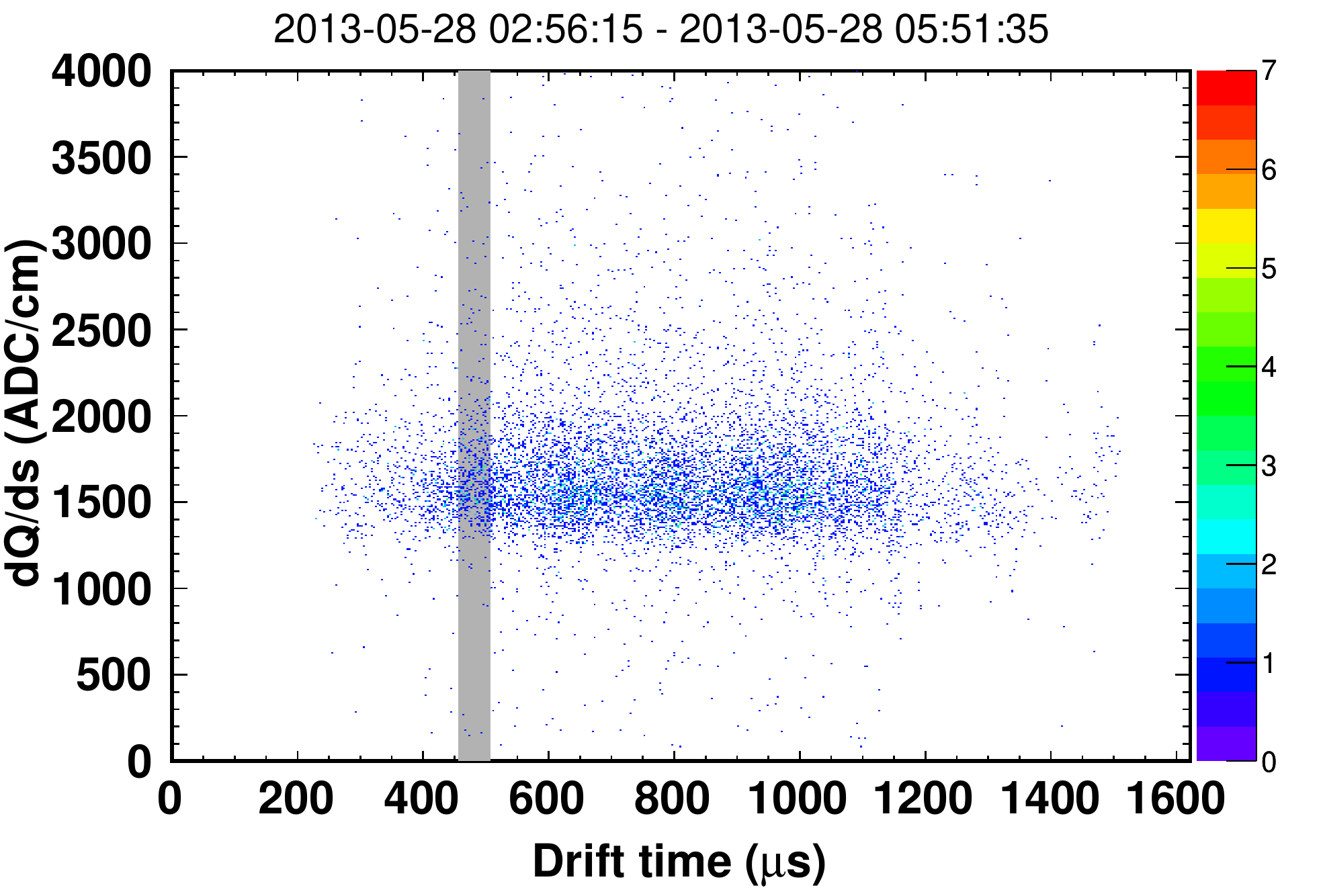}
\includegraphics[width=.45\textwidth]{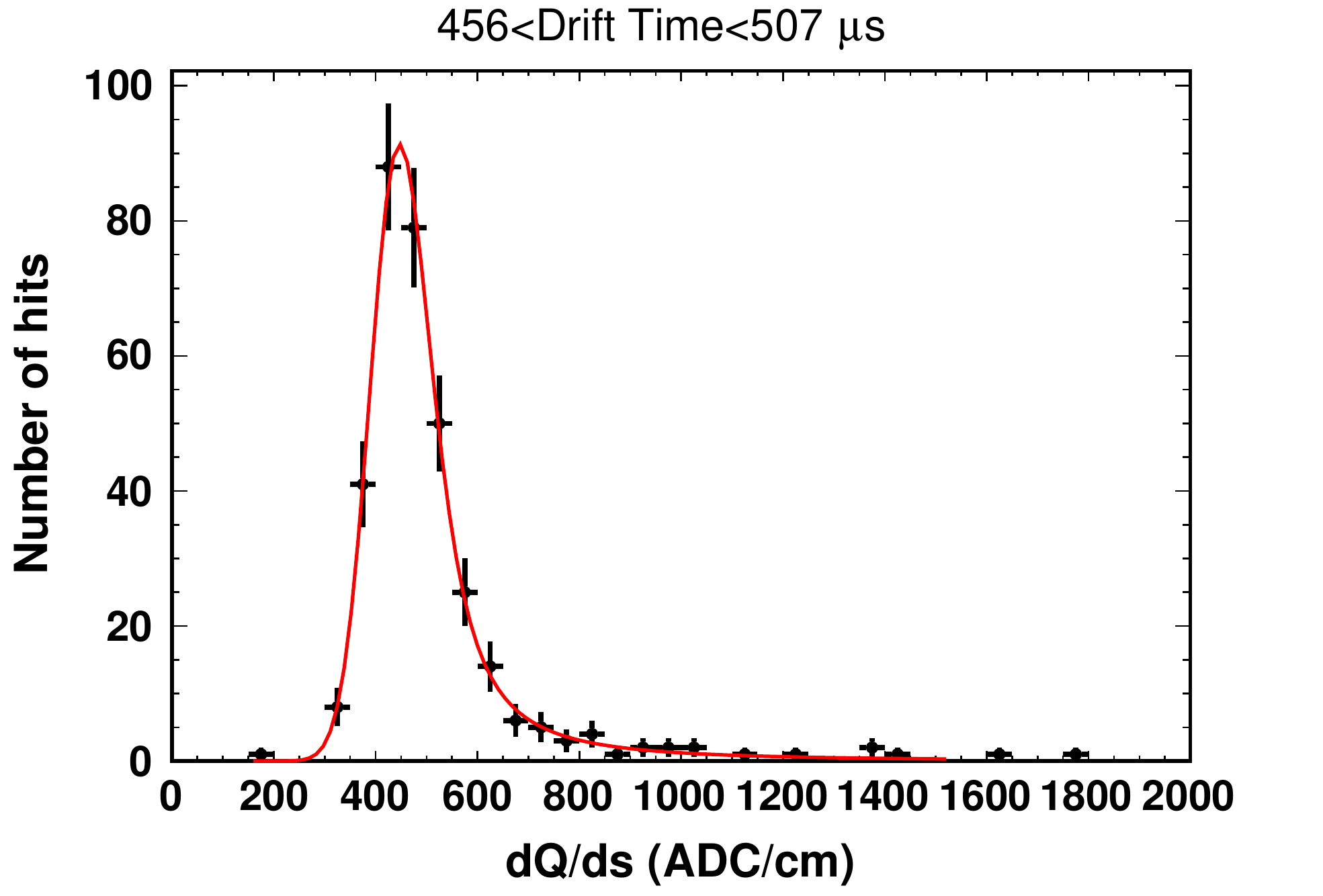}
\includegraphics[width=.45\textwidth]{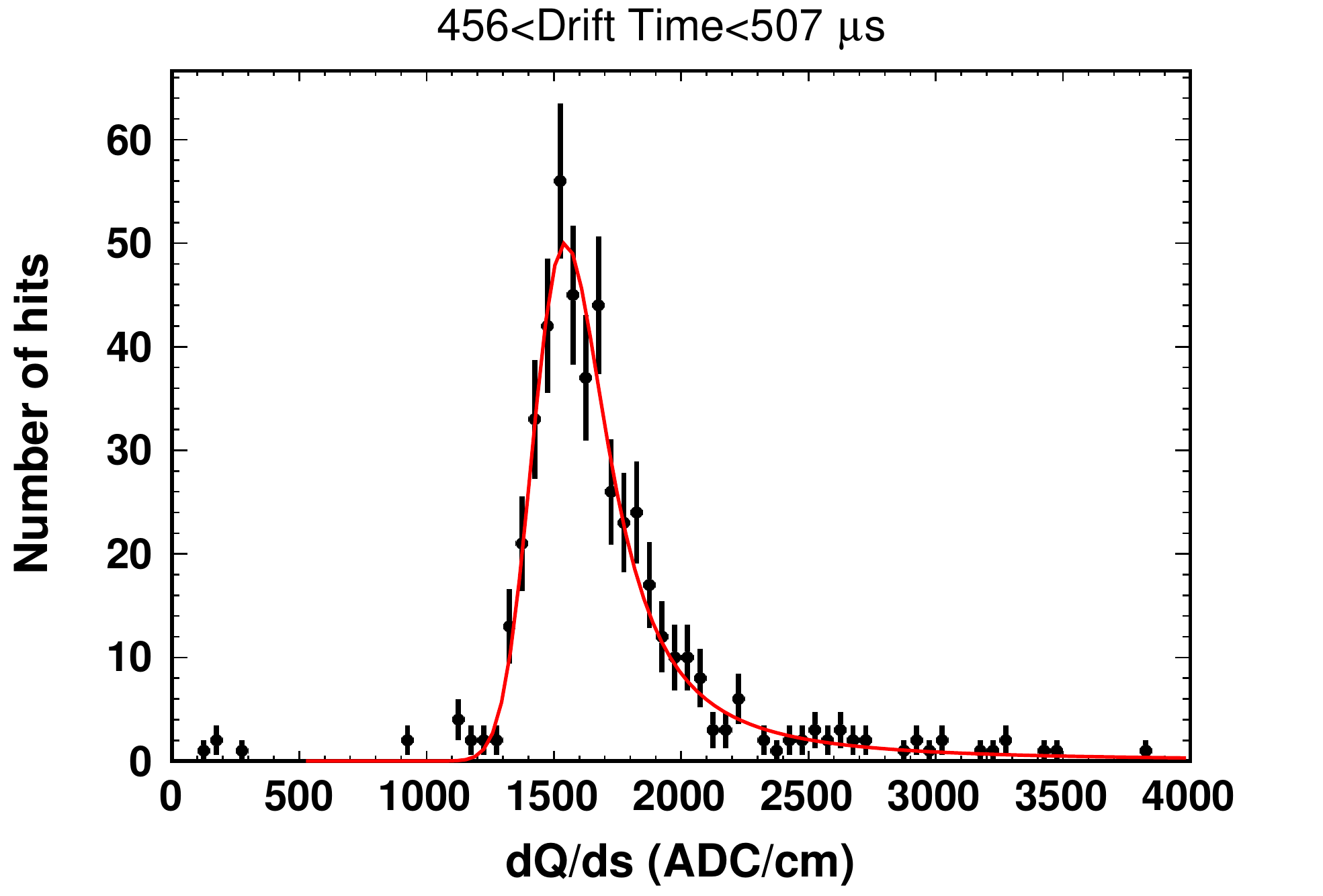}
\includegraphics[width=.45\textwidth]{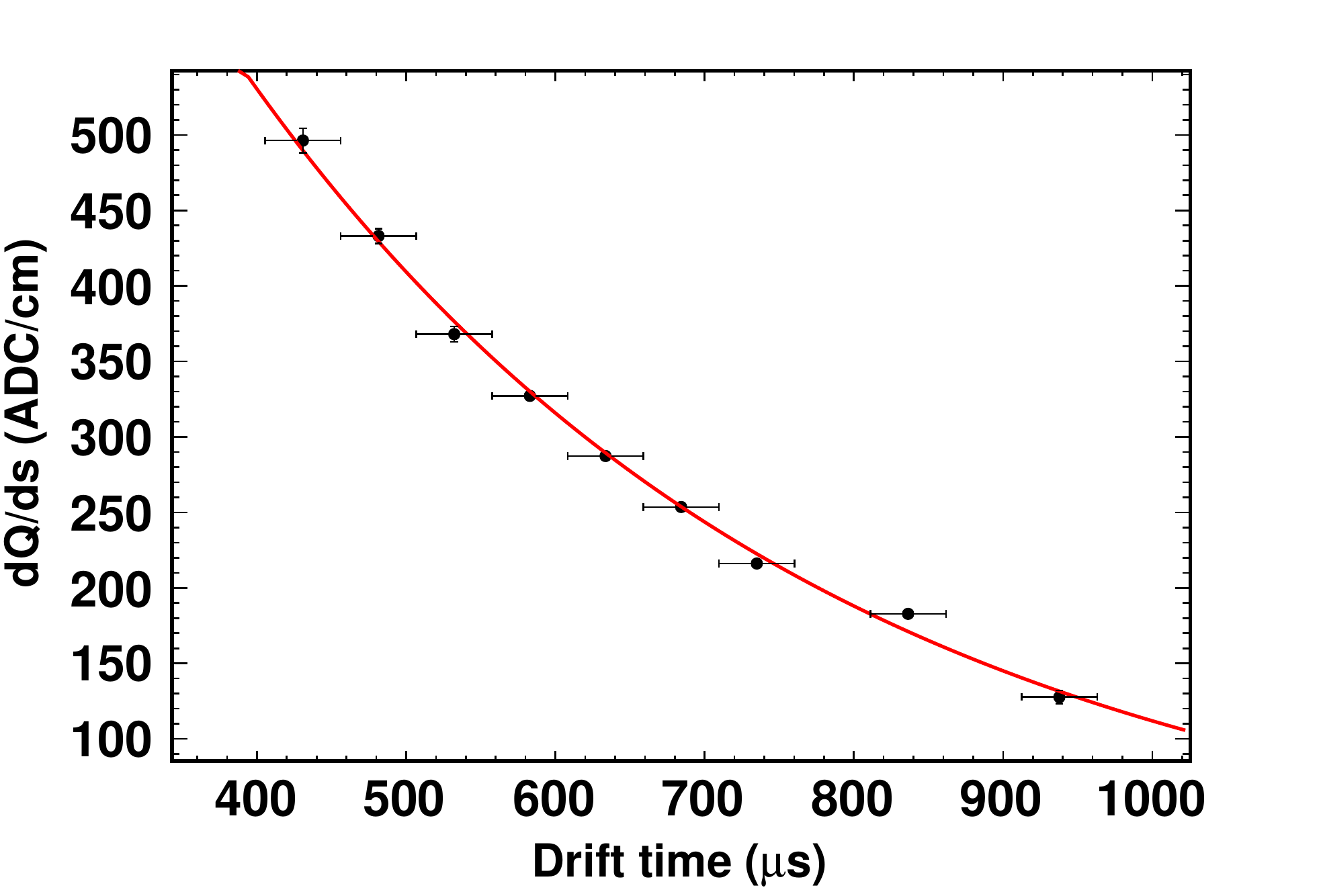}
\includegraphics[width=.45\textwidth]{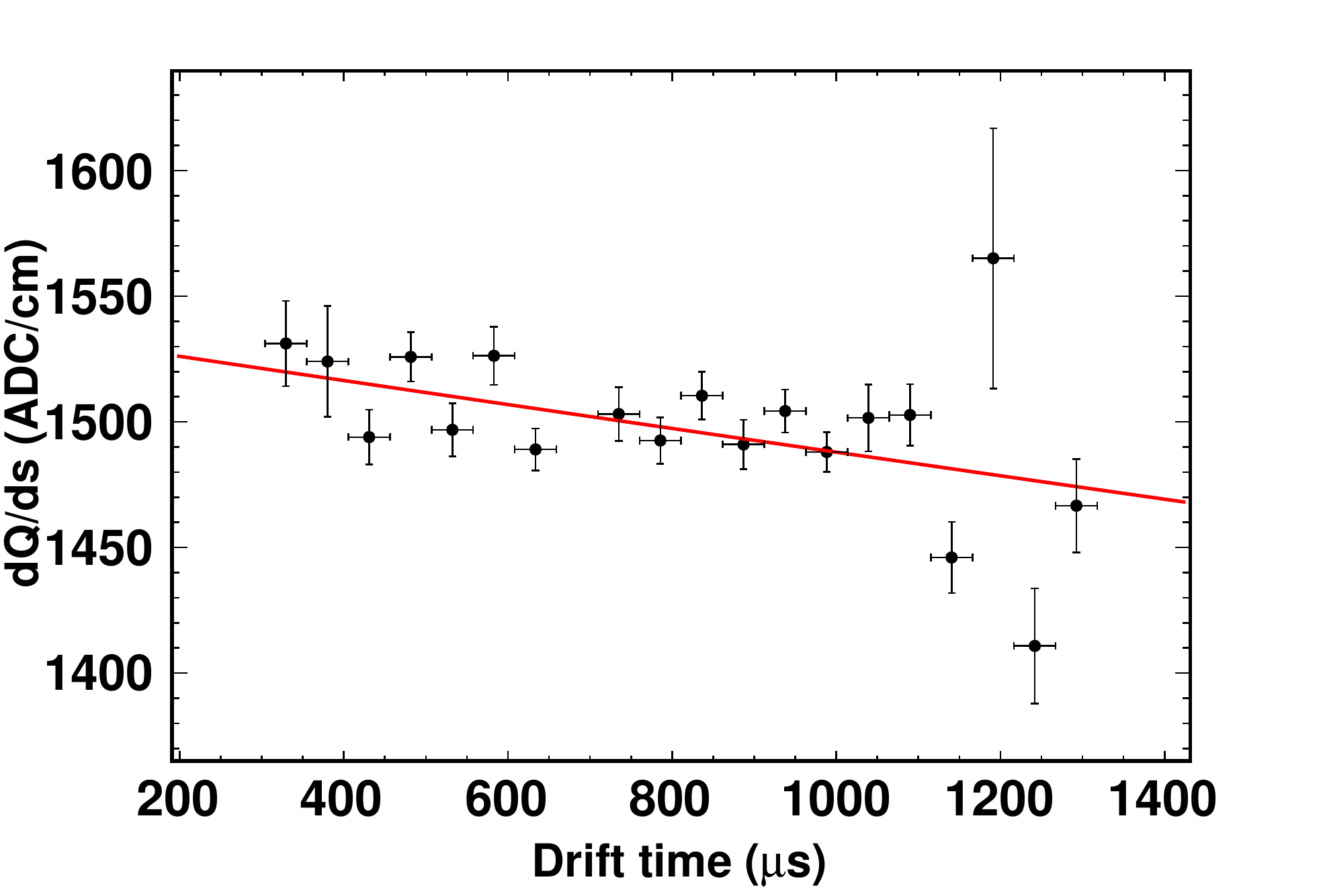}
\caption{$dQ/ds$ distributions of hits using data taken in two periods of time when the electron lifetime was 0.4 ms (left) and 30 ms (right), respectively. Top: Scatter plots of $dQ/ds$ as a function of electron drift time; Middle: $dQ/ds$ distributions for hits with drift time between 456 and 482 $\mu s$; Bottom: $dQ/ds$ from Landau fit as a function of drift time.}
\label{fig:landau} 
\end{figure}

Figure~\ref{fig:landau} shows the hit $dQ/ds$ distributions for a calculated electron drift lifetime of 0.4 ms (left) and of 30 ms (right). The top plots show $dQ/ds$ versus electron drift time distributions. The middle plots show the $dQ/ds$ distributions for hits with drift time betwen 456 and 507 $\mu s$, indicated as gray bands in the top plots, as well as the Landau convoluted with Gaussian fits. The bottom plots shows the exponential fits.

\begin{figure}[htbp]
  \begin{center}
    \includegraphics[width=1\textwidth]{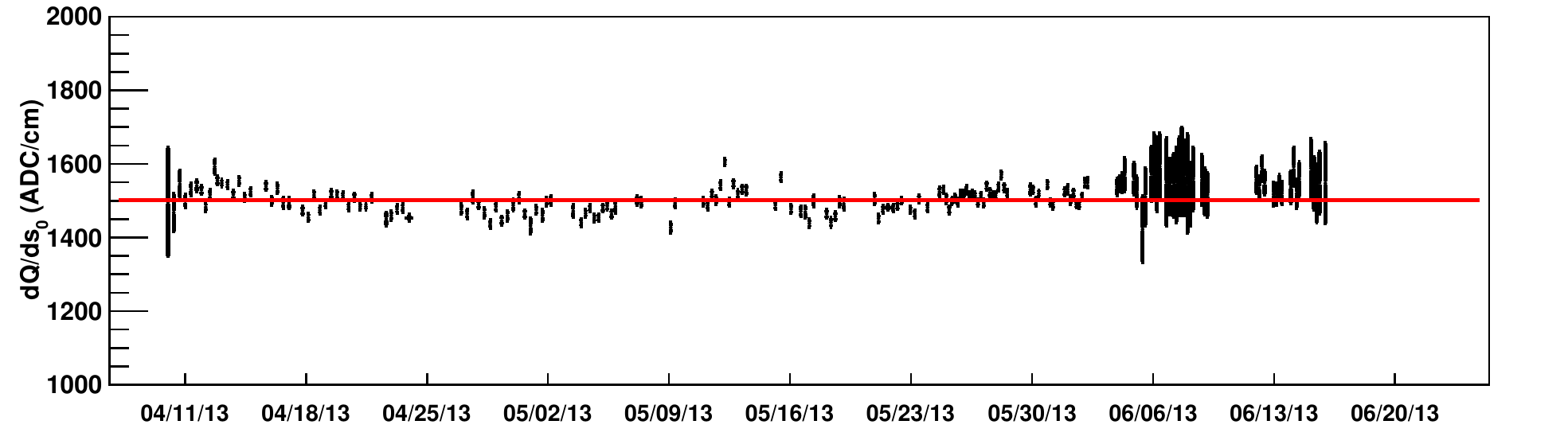}
    \caption{$dQ/ds_{0}$ as a function of time.}
\label{fig:dqdx0}
\end{center}
\end{figure}

As a measure of the stability of the entire system, we plot in figure~\ref{fig:dqdx0} the calculated $dQ/ds_{0}$ over the period of a two-month run. This quantity stays remarkably constant.

\begin{figure}[htbp]
  \begin{center}
    \includegraphics[width=1\textwidth]{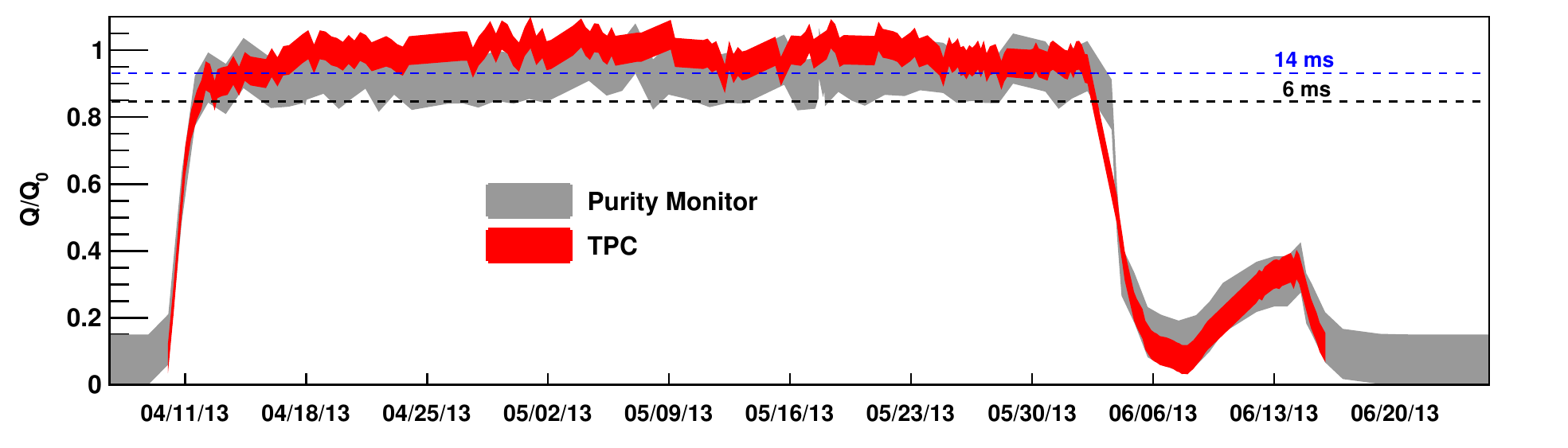}
    \caption{Comparison between $Q/Q_{0}$ measured by the TPC and the converted $Q_/Q_{0}$ using the purity monitor data. $Q/Q_{0}$ is a measure of the electron lifetime, and the values $\tau = $ 6 ms and $\tau = $ 14 ms that correspond to $Q/Q_{0}$ = 0.85 and $Q/Q_{0}$ = 0.93, respectively, are indicated with dashed lines. 
}
\label{fig:qaqc}
\end{center}
\end{figure}

It is interesting to compare the measurement from the TPC with the purity monitor. Figure~\ref{fig:qaqc} shows the charge ratio for a 1 ms $\Delta t_{drift}$
\begin{equation}
Q/Q_{0} = e^{-\Delta t_{drift}/\tau} = e^{-a\Delta t_{drift}}
\end{equation}
as a function of calendar time. We use $\Delta t_{drift} = 1$ ms, which is the typical time range in the TPC measurement. To make the comparison, we convert the anode-to-cathode ratio ($Q_{A}/Q_{C}$) measurement from the purity monitor for a 0.31 ms drift time~\cite{Adamowski:2014daa} to the $Q/Q_{0}$ one would obtain for a 1 ms drift time\footnote{The sources of systematic uncertainty on the purity monitor measurement are discussed in Ref.~\cite{Adamowski:2014daa} and shown on the plot as a combined 15\% absolute uncertainty on the converted $Q/Q_{0}$.}. 


\begin{figure}[htbp]
  \begin{center}
    \includegraphics[width=0.8\textwidth]{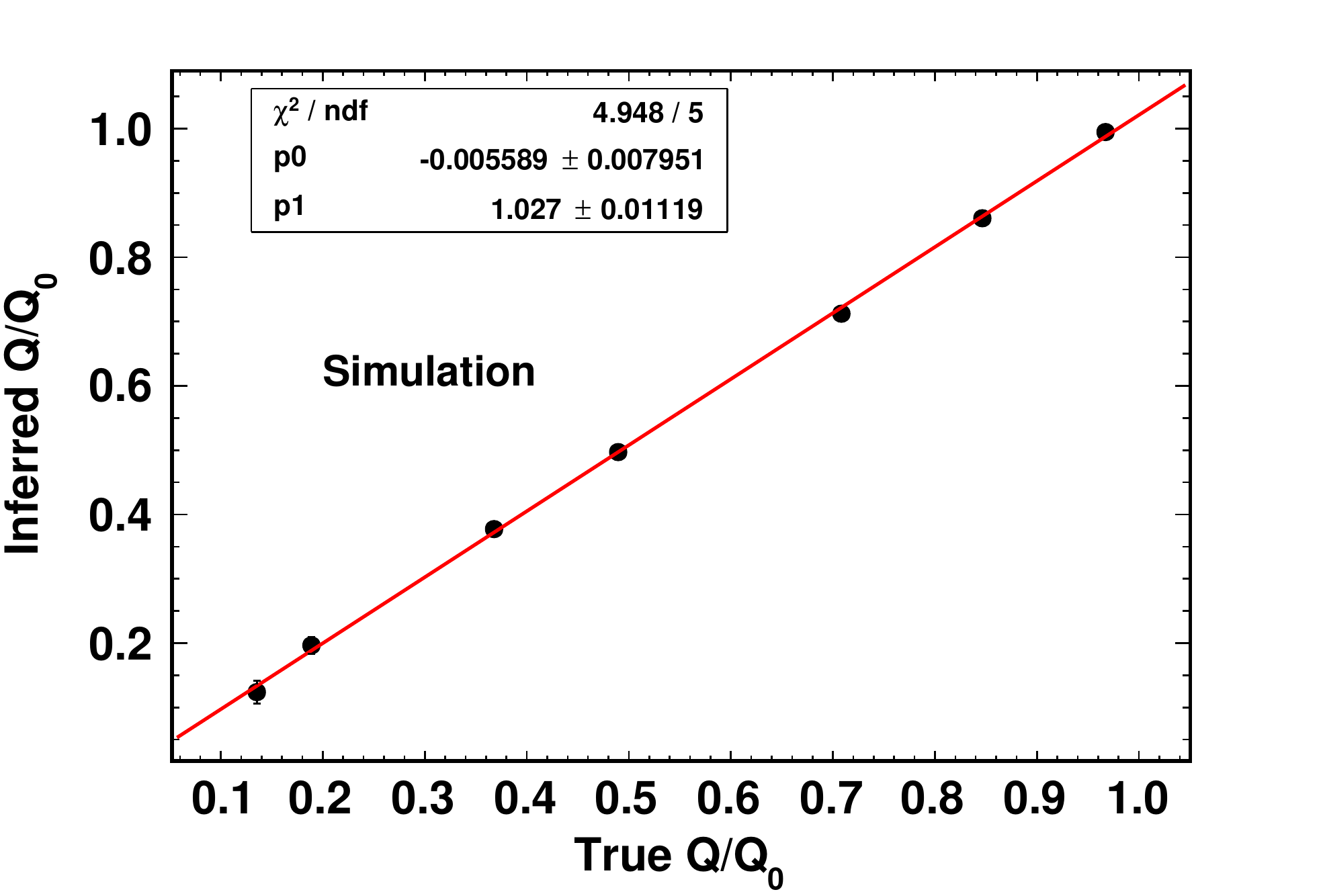}
    \caption{$Q/Q_{0}$ determined using Monte Carlo simulated events and the method described in the text compared to the value of $Q/Q_{0}$ used in the simulation.
}
\label{fig:simqaqc}
\end{center}
\end{figure}

In order to evaluate the robustness of the lifetime measurement and the extent of the systematic uncertainties, we test the same method using a sample of 100 000  simulated cosmic ray muon tracks in the LongBo TPC. The simulation was done with {\sc larsoft}~\cite{larsoft} and includes the muon ionization and recombination effects, the free electron longitudinal and transverse diffusion, the signal attenuation due to the impurities in liquid argon and the electronics response. The method has been tested for electron lifetimes in the range of 0.5 - 30 ms. The results are summarized in figure~\ref{fig:simqaqc}, which compares the $Q/Q_0$ input and the $Q/Q_0$ inferred. The measurement technique is reliable in a large range of liquid argon lifetime values. We fit the points to a straight line. The linear fit result
\begin{equation}
(Q/Q_{0})^{Inferred} = -0.006 + 1.027\times(Q/Q_{0})^{True}
\end{equation}
is used to correct the data measurements to obtain the results shown in figure~\ref{fig:qaqc}.

We consider two systematic effects in the purity measurement: field non-uniformity and system instability, which are described as follows. It is possible that the electric field in the TPC is not completely uniform because of the presence of the positive ions (space charge) and other effects. However, we have evidence that the field uniformity is fairly well preserved because \emph{i}) the maximum drift time we observe in the TPC data is consistent with the drift time for an electric field of 350 V/cm over 2 m and \emph{ii}) we do not observe distortion of tracks.

The method we employ to measure the electron drift lifetime is robust against random non-uniformity in the electric field. The measurement depends on the electron drift time, which is measured by the scintillation counters and the data acquisition system to a good precision (less than 1 $\mu s$) and is unaffected by the non-uniform electric field. We do not use signal recorded on the boundary wires to avoid edge effects near the field cage. 

There is a 1\% uncertainty in the high voltage applied to the cathode resulting from the power supply monitor and filter resistors. We varied the magnitude of the constant electric field by $\pm5\%$ conservatively in the reconstruction code and the resulting difference in the $Q/Q_{0}$ is negligible. 

The accumulation of space charge in the TPC can change the electric field in a way that affects the electron drift lifetime measurement directly. Ref.~\cite{Palestini:1998an} predicts the electric field is increasing at longer drift distance as $\propto\sqrt{1+\beta x^2}$ where $x$ is the drift distance with respect to the anode and $\beta$ is a constant that depends on the chamber length, the applied high voltage and the cosmic ray rate. Because a larger electric field suppresses the electron-ion recombination, one expects more charge at longer drift distance, and the measured signal attenuation is smaller than the real attenuation in the presence of a space charge induced electric field distortion. We introduced a non-uniform field of this form in the simulation such that the field at the cathode is 20\% larger than the field at the anode. The resulting $Q/Q_{0}$ is shifted by 0.03.  Because we do not know what fraction of the argon ions were absorbed by the field cage before arriving at the cathode, we take 0.03 as the systematic uncertainty from undetected non-uniformity in the drift field.

We also study the system instability by looking at the distribution of $Q/Q_{0}$ when the value is consistent with 1 (left of plot in figure~\ref{fig:qaqc}). The distribution shows a mean value of 0.99 with a width of 0.03. We take 0.03 as systematic uncertainty from system instability. The combined systematic uncertainty is 0.04 and is shown in figure~\ref{fig:qaqc}. Following the Bayesian approach~\cite{pdg} we set a lower limit on $Q/Q_{0}$ to be 0.93 at 95\% confidence level, which corresponds to a lifetime of 14 ms.

\subsection{High Voltage Stability}
\begin{figure}[htb]
  \centering
  \includegraphics[width=\textwidth]{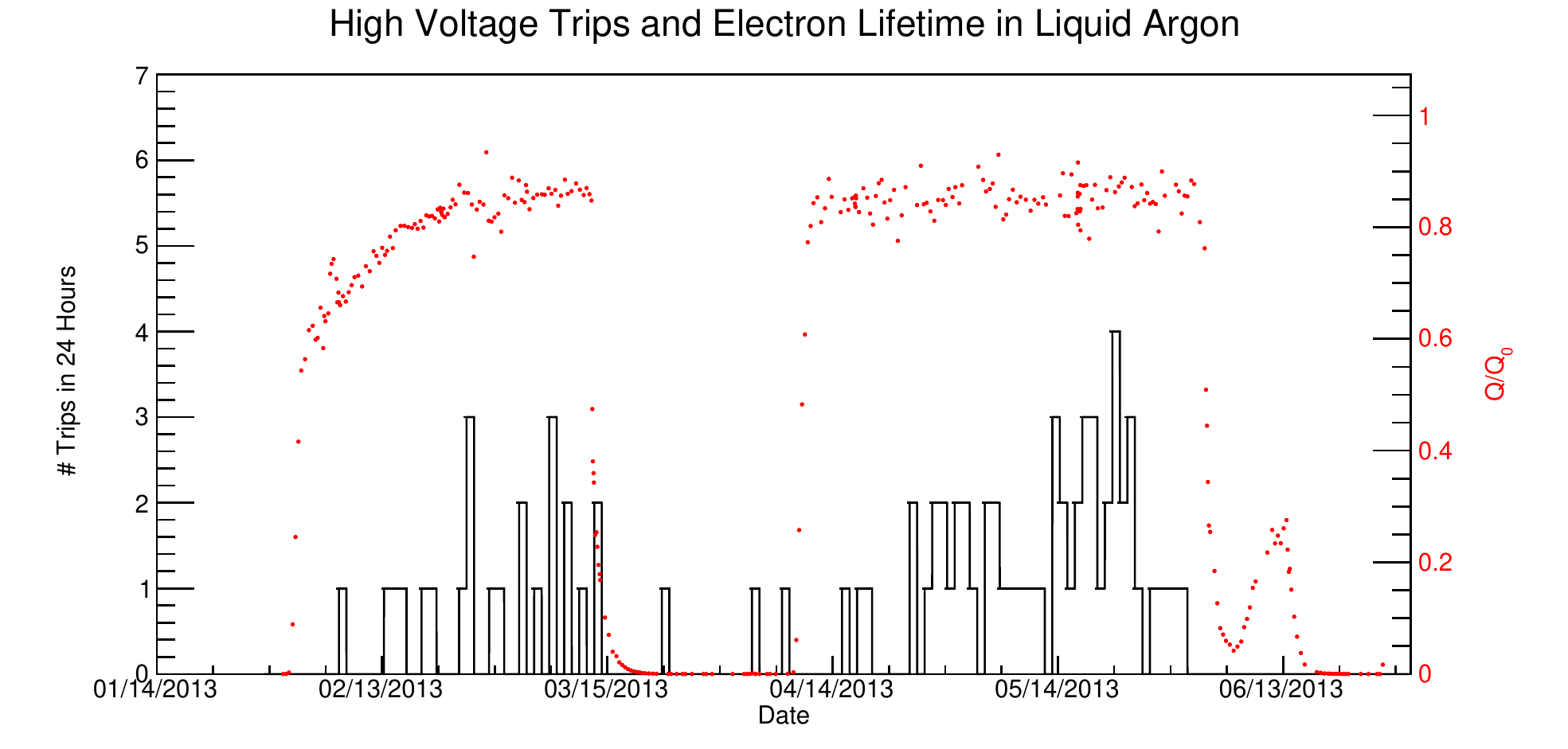}
  \caption{The number of cathode high voltage trips in a 24~hour period and purity monitor values during data taking.  The trip counts are given by the black solid line and scale on the left.  Trips related to testing the high voltage system were not included in the trip count.  A quantification of the argon purity is given by the red dots and the scale on the right, which is the converted $Q/Q_0$ using purity monitor data.}
  \label{fig:voltageVsPurity}
\end{figure}

The power supply was set to turn off if it drew an over-current much greater than the operating current of the TPC.  During data taking, the cathode high voltage never sustained greater than 80~kV for a significant period with most data being collected with 60 or 70~kV on the cathode.  A correlation was noted between greater high voltage stability and greater electronegative contamination in the argon. This trend can be seen in figure~\ref{fig:voltageVsPurity} where the number of high voltage trips is displayed along with a purity monitor reading as a function of time.  While a precise characterization of the dependence is beyond the scope of this paper, the behavior agrees with other studies in the literature that have shown a trend between an increase in electronegative contamination and an increase in liquid argon's dielectric strength \cite{ref:swan, ref:bern, ref:uBLarBreakdown}.

\section{Conclusions}\label{sec:conclusions}

In this paper we have discussed the design and operation of the LongBo TPC 
in the LAPD cryostat.  A ratio $S/N=50.5/1.69\sim30$ was measured for the wires in the collection
plane with discrete CMOS electronics.
For these data the TPC was operated with a drift field of 350~V/cm.
The measured $S/N$ ratio for the ASIC channels was 1.4 times larger
than that measured for the discrete channels for the ASIC shaping time
of 2.0~$\mu$s and a gain setting of 25 mV/fC. The liquid argon purity measurement using cosmic
ray muons in the TPC shows the electron drift lifetime is at least 14 ms at 95\% confidence level over a long period of time.
We also observe a correlation between the voltage stability and the liquid
argon purity.

\acknowledgments
We would like to thank Tom Junk for the useful discussions on statistical analysis. This work was supported by the US DOE, and by the NSF through Grants 1068318 and 1410972, to Michigan State University. We thank the staff at Fermilab for their technical assistance in running the LAPD experiment. Fermilab is operated by Fermi Research Alliance, LLC under Contract No. De-AC02-07CH11359 with the United States Department of Energy.


\begin{thebibliography}{9}
\bibitem{Amerio:2004ze} 
  S.~Amerio et al. (ICARUS Collaboration),
  \emph{Design, construction and tests of the ICARUS T600 detector},
  {\emph{Nucl.\ Instrum.\ Meth.\ A} {\bf 527}, (2004) 329}.

\bibitem{Anderson:2012vc} 
  C.~Anderson et al. (ArgoNeuT Collaboration),
  \emph{The ArgoNeuT detector in the NuMI low-energy beam line at Fermilab},
  \jinst{7}{2012}{P10019}.

\bibitem{Ereditato:2013xaa} 
  A.~Ereditato, {\it et al.},
  \emph{Design and operation of ARGONTUBE: a 5 m long drift liquid argon TPC},
  \jinst{8}{2013}{P07002}.


\bibitem{Adamowski:2014daa} 
  M.~Adamowski et al.,
  \emph{The Liquid Argon Purity Demonstrator}, 
\jinst{9}{2014}{P07005}.

\bibitem{bo}
C.~Bromberg, \emph{First neutrino physics results of the ArgoNeuT experiment, and first tracks with CMOS preamplifiers operating at 87 K}, in Proceedings of the 2nd International Workshop towards the Giant Liquid Argon Charge Imaging Experiment (GLA2011), 5-10 Jun 2011, Jyv\"askyl\"a, Finland, lartpc-docdb.fnal.gov/0008/000819/001/CBromberg\_GLA2011\_Proc.pdf, to be published.

\bibitem{ohmite}
Ohmite model 104, rated at 10 kV, 1 W.

\bibitem{ref:glassman}
Glassman High Voltage Inc., PO Box 317, 124 West Main Street, High Bridge, NJ  08829-0317, U.S.A.
\href{http://www.glassmanhv.com}{www.glassmanhv.com}.

\bibitem{Bromberg:2001gd} 
  C.~Bromberg,
  \emph{Gain and threshold control of scintillation counters in the CDF muon upgrade for Run II},
   {\emph{Int.\ J.\ Mod.\ Phys.\ A} {\bf 16S1C} (2001) 1143}.

\bibitem{Antonello:2014eha} 
  M.~Antonello  et al.,
  \emph{Experimental observation of an extremely high electron lifetime with the ICARUS-T600 LAr-TPC},
\jinst{9}{2014}{P12006}.

\bibitem{Palestini:1998an} 
  S.~Palestini et al.,
  \emph{Space charge in ionization detectors and the NA48 electromagnetic calorimeter},
  {\emph{Nucl.\ Instrum.\ Meth.\ A} {\bf 421} 1999 75}.

\bibitem{ref:swan}
D.~Swan and T.~Lewis, \emph{Influence of electrode surface conditions on the electrical strength of liquified gases}, {\emph{J.\ Electrochem.\ Soc.} {\bf 107} (1960) 180}.

\bibitem{ref:bern}
A.~Blatter et~al., \emph{Experimental study of electric breakdowns in liquid
  argon at centimeter scale},
  \jinst{9}{2014}{P04006}.

\bibitem{ref:uBLarBreakdown} 
  R.~Acciarri et al.,
  \emph{Liquid argon dielectric breakdown studies with the MicroBooNE purification system},
  \jinst{9}{2014}{P11001}.

\bibitem{Thorn:2012zsa} 
  C.~Thorn et al.,
  \emph{Cold Electronics development for the LBNE LAr TPC},
  {\emph{Phys.\ Procedia} {\bf 37} (2012) 1295}.

\bibitem{Adams:2013qkq} 
  C.~Adams et al.  (LBNE Collaboration),
  \emph{The Long-Baseline Neutrino Experiment: exploring fundamental symmetries of the universe}, 
{\href{http://arxiv.org/abs/1307.7335}{\tt hep-ex/1307.7335}}.

\bibitem{microboone}
\href{http://www-microboone.fnal.gov/publications/TDRCD3.pdf}{http://www-microboone.fnal.gov/publications/TDRCD3.pdf}.


\bibitem{larsoft}
\href{https://cdcvs.fnal.gov/redmine/projects/larsoft}{https://cdcvs.fnal.gov/redmine/projects/larsoft}. We use version v02\_00\_01 of this software package.

\bibitem{pdg}
K.A.~Olive et al. (Particle Data Group), {\emph{Chin.\ Phys.\ C} {\bf 38} (2014) 090001}.


\end{thebibliography}
\end{document}